\documentclass[preprint,review,12pt]{elsarticle}
\usepackage{graphicx,stfloats}
\usepackage{color}
\usepackage[version=4]{mhchem}

\usepackage{amsmath}
\usepackage{hyperref}
\usepackage{longtable}
\usepackage{subcaption}
\usepackage{algorithm}
\usepackage{algpseudocode}
\usepackage{soul}
\usepackage{tikz}
\usetikzlibrary{patterns}
\newcommand{\del}{\vec{\nabla}}

\usepackage{subcaption}
\begin{document}
\begin{frontmatter}
                                    
\title{Vidyut3d: a GPU accelerated fluid solver for non-equilibrium plasmas on adaptive grids.}

\author[nrel]{Hariswaran Sitaraman\corref{cor1}}
\ead{hariswaran.sitaraman@nrel.gov}

\author[nrel]{Nicholas Deak}

\author[minnesota]{Taaresh Taneja}

\address[nrel]{Computational Science Center, National Renewable Energy Laboratory, 15013 Denver West Parkway, Golden, CO, 80401, USA}

\address[minnesota]{Department of Mechanical Engineering, University of Minnesota -- Twin Cities, Minneapolis, MN 55455, USA}

\cortext[cor1]{Corresponding author:}

\begin{abstract}
We present the numerical methods, programming methodology, verification, and performance assessment of a non-equilibrium plasma fluid solver that can effectively utilize current and upcoming central processing and graphics processing unit (CPU+GPU) architectures, in this work. Our plasma fluid model solves the coupled conservation equations for species transport, electrostatic Poisson and electron temperature on adaptive Cartesian grids. Our solver is written using performance portable adaptive-grid/particle management library, AMReX, and is portable over widely available vendor specific GPU architectures. We present verification of our solver using method of manufactured solutions that indicate formal second order accuracy with central diffusion and fifth-order weighted-essentially-non-oscillatory (WENO) advection scheme. We also verify our solver with published literature on capacitive discharges and atmospheric pressure streamer propagation. We demonstrate the use of our solver on two 3D  simulation cases: an atmospheric streamer propagation in \ce{Ar}-\ce{H2} mixtures and a low pressure twin electrode radio frequency reactor. Our performance studies on three different CPU+GPU architectures indicate  $\sim$ 150-400X speed-up using AMD and NVIDIA GPUs per time step compared to a single CPU core for a 4 million cell simulation with 15 species.
\end{abstract}


\end{frontmatter}
\ifdefined \wordcount
\clearpage
\fi
\section{Introduction}
\label{sec:intro}
Many engineering applications such as semiconductor manufacturing, materials/chemical synthesis and surface modifications \cite{lieberman1994principles} depend on non equilibrium or non-thermal plasmas. These plasma discharges are characterized by a property wherein the electrons, energized by input electrical power, approach  temperatures $\sim$ 10,000-100,000 K, while the gas remains at much lower temperatures (300-1000 K). 
These high energy electrons can therefore enable reactive chemistry at normal temperatures and pressures and are actively investigated in recent times for plasma-assisted ignition \cite{blankson2016,leonov2006plasma,lefkowitz2018reduction},  chemicals manufacturing from stable molecules (e.g., \ce{CO2}, \ce{N2}, \ce{H2}) \cite{delikonstantis2022low},  metal refining \cite{satritama2024hydrogen}, actuation in supersonic flows \cite{starikovskii2009sdbd}, and liquid treatment  \cite{bruggeman2016plasma,foster2018towards,kruszelnicki2019atmospheric}, among others. 
The simulation of plasma phenomena that arise in many of these applications (e.g. propagating streamer heads, and plasma sheaths) is a challenging problem characterized by a wide range of spatial and temporal scales, and tight coupling among physical processes. A computational framework for solving the plasma governing equations is essential for understanding the interactions among complex electron-impact reactions, gas-phase/surface chemistry and diffusive/mobility driven species transport. Such a predictive framework will enable improved designs, control of plasma interaction processes, and energy efficient plasma generation techniques. 

Non-equilibrium plasmas have been simulated  mainly through fluid and particle based approaches \cite{kim2005particle}. Particle based methods using Particle-in-Cell-Monte-Carlo-collisions (PIC-MCC) are used in the low pressure regime ($\sim$ several mTorrs to 1 Torr) where number densities and collisional effects are lower compared to high pressure regimes. When the number densities become higher, for example at atmospheric conditions, a fluid model can accurately describe plasma chemistry and dynamics, with reduced costs compared to using large number of super-particles (several millions or more) in a PIC-MCC simulations \cite{kim2005particle}. 

There are a number of existing codes that have been developed to simulate non-equilibrium plasmas, including MOOSE \cite{lindsay2016fully}, HPEM \cite{kushner2009hybrid}, COMSOL Multiphysics, PLASIMO \cite{van2009plasma}, and SOMAFOAM \cite{verma2021somafoam}. There are, however, limitations with these existing codes, as they are not all open-source and/or do not support hybrid CPU+GPU parallelization support, limiting their ability to take advantage of modern HPC architectures. There have been several recent efforts on developing large scale particle codes for non-equilibrium plasmas that can utilize current heterogeneous high performance computing architectures \cite{vay2018warp,almgren2022multi}. The development of performance portable programming paradigms (e.g., KOKKOS \cite{trott2021kokkos}) has enabled solvers to utilize both central processing and graphics processing units (CPU+GPU) concurrently. On the other hand, non-equilibrium plasma fluid solvers present additional complications due to several coupled linear solves, electron equation -Poisson coupling, and meshing requirements for resolving relevant length scales in high pressure discharges, that make it difficult for GPU portability and performance. To date, there are no open-source performance-portable non-thermal plasma fluid solvers capable of running efficiently on CPU+GPU architectures. 

The main contribution of this work is therefore aimed towards the development of a performance portable fluid solver for non-equilibrium discharges. We present the mathematical model (section \ref{sec:goveq}), numerical methods (section \ref{sec:numerics}), software implementation (section \ref{sec:implementation}) for such a solver, verify the code (section \ref{sec:codeverify}), demonstrate its use in production simulations (section \ref{sec:demo}) and show performance implications on CPU+GPU architectures (section \ref{sec:perf}) in this work.

\section{Mathematical model}
\label{sec:goveq}
\subsection{Governing equations}
A two temperature model that is well established for modeling non-equilibrium discharges \cite{deconinck2007,sitaraman2011,sitaraman2012,breden2012} at intermediate and high pressure regimes is implemented in this work. The model mainly consists of conservation equations for electrons, ions and neutral species, an electrostatic Poisson equation to obtain self-consistent electrostatic potential and electric fields, the drift-diffusion approximation for species fluxes and the electron energy equation to find electron temperature. These equations are as described below. Our model currently does not solve the conservation equations for the background gas; we assume constant pressure and temperature for the dominant background species, which is typically the case for non-equilibrium weakly ionized plasmas. The coupling of our plasma model with fluid equations for the background gas is part of our ongoing efforts.

The conservation equation for the number densities of each species ($n_k$) 
including electron, ions and neutral species are given by:
\begin{equation}
\label{continuity}
\frac{\partial{n_k}}{\partial t} + \del \cdot {\vec{\Gamma}_k}=\dot{G}_k
\end{equation}
$\dot{G}_k$ denote the generation/loss term for species $k$ through chemical reactions in the plasma. The flux term $\vec{\Gamma}_k$ is given by the drift-diffusion
approximation:
\begin{equation}
	\vec{\Gamma}_k = n_k \vec{u}_k = \mu_k n_k \vec{E} - D_k \del n_k,
	\label{driftdiff}
\end{equation}
Here $\mu_k$ and $D_k$ denote species mobility and diffusion coefficient, respectively. 
The species flux $\vec{\Gamma}_k$ reduces to 
Fick's law of diffusion for neutral species that have zero mobility. $\vec{E}$ is the self-consistent electric field obtained from
solving the electrostatic Poisson equation:
\begin{eqnarray}
	\nabla^2 \phi + \frac{e}{\epsilon_0}{\sum_{k=1}^{N} Z_k n_k}  = 0, \label{eq:poisson} \\
	\nonumber \\
	\vec{E}=-\del \phi
	\label{eq:efield}
\end{eqnarray}
$\phi$ represents the electrostatic potential and the summation term represents
the net space charge per unit volume. Here $Z_k$ represents the species charge (e.g., +1 for
singly positive ions, -1 for electrons, 0 for neutrals) and the summation is performed
over total number of species ($N$). The electronic
charge and permittivity of free space are represented by $e$ and $\epsilon_0$, respectively.

The species production terms for electron impact reactions, mobilities 
and diffusion coefficients depend on the local electron temperature which is obtained
by solving the electron energy equation, given by:
\begin{equation}
	\frac{\partial {E_e}}{\partial t} + \del \cdot \vec{\Gamma}_{\epsilon}= \dot{S}_{\epsilon}
	\label{eenergy_eq}
\end{equation}
Here $E_e$ denotes electron energy, given by:
\begin{equation}
	E_e=\frac{3}{2} n_e k_B T_e
\end{equation}
$T_e$ represents the electron temperature and $n_e$ is electron number density. $\Gamma_{\epsilon}$ is the electron energy flux which is obtained from the drift diffusion approximation for electron energy as:
\begin{equation}
\Gamma_{\epsilon}= \frac{5}{3} \mu_e E_{e} \vec{E} - \frac{5}{3} D_e \del E_{e}   
\label{driftdiff_een_eq}
\end{equation}
The source term on the right hand side of Eq.\ \ref{eenergy_eq} is given by
\begin{equation}
	\dot{S}_e = -e\vec{\Gamma}_e \cdot \vec{E} - \frac{3}{2}n_e k_B (T_e - T_g) \frac{2 m_e}{m_b} \nu
	- \sum_i \Delta E_i r_i
	\label{eenergysource}
\end{equation}
Here, $T_g$ represents the gas temperature, $m_e$ is the mass of electron and $m_b$ is the mass of the dominant background neutral
species. $\nu$ represents the electron-background neutral collision frequency, which can be obtained from electron mobility using
expression, $\mu_e = \frac{e}{m_e \nu}$ or from elastic collision rate computed from offline Boltzmann solves. 
The first term on the RHS here represents the electron Joule heating term, which is the energy gained by the
electrons from applied and induced electric fields. The second term is the volumetric energy loss due to elastic collisions of 
electrons with the background neutral species. The final term is the effect of plasma chemistry and 
represents net energy loss due to electron inelastic collisions, mainly 
via electron-impact ionization/excitation reactions. Here, $\Delta E_i$ is the energy associated 
with reaction $i$ that involves electrons and $r_i$ is the rate of progress of that particular reaction.

Our model also can incorporate photoionization source term for electrons and ions, ${S}_{ph}$ in units of $\mathrm{m^{-3}s^{-1}}$, that can be included for relevant simulations. In the context of streamer discharges, photoionization aids in the creation of seed electron clouds near positively charged streamer head, resulting in secondary avalanches that enhance streamer propagation. 
%
Zhelznyak et al.\ \cite{zhelezniak1982photoionization} who developed the original model for photoionization, proposed a Green's function based evaluation of the photoionization source term, where photoionization at every point in the domain was due to the emission of photons at every point in the domain. Given the computational expense of this Green's function based evaluation in 2D and 3D configurations, we use Bourdon et al.'s~\cite{bourdon2007efficient} simplified three-term Helmholtz equation based solution, (Eq.~\ref{eq:photoion}), where the coefficients $A_j$ and $\lambda_j$ for $j=1,2,3$ depend on specific gas mixtures, that can be obtained by fitting the analytical absorption function using sum of exponential terms, as is done in~\cite{bourdon2007efficient}. Values of these coefficients for a pure air discharge at 300 K have been provided in Bourdon et al.~\cite{bourdon2007efficient}. 
\begin{eqnarray}
\label{eq:photoion}
 \nabla^2{S_{ph}^{j}}(\vec{r}) - (\lambda_j p_{\ce{O2}})^2{S_{ph}^{j}}(\vec{r})  = -A_j p_{g}^2 I(\vec{r})\,\, j=1,2,3 \\
 S_{ph}=S_{ph}^1+S_{ph}^2+S_{ph}^3
\end{eqnarray}

In our solver, we have the option of using the definition of the emission function as is used in Breden et al.~\cite{breden2013aip}, where $I\vec{(r)}$ is assumed to be proportional to the local ionization rate of the dominant neutral species (e.g., \ce{N2} in an air plasma), $S_i(\vec{r})$ and is given in Eq.\ \ref{eq:emission}, or that proposed in Bourdon et al.~\cite{bourdon2007efficient}, where it is also assumed to be proportional to the ratio of the electronic excitation rate ($\nu_u$) to the ionization rate ($\nu_i$) of background neutral species (Eq.\ \ref{eq:emission2}).
\begin{equation}
\label{eq:emission}
I(\vec{r}) = \frac{p_q}{p + p_q}\xi S_i(\vec{r}).
\end{equation}

\begin{equation}
\label{eq:emission2}
I(\vec{r})= \frac{p_q}{p + p_q} \xi \frac{\nu_u}{\nu_i} S_i(\vec{r})
\end{equation}


%
We also account for surface charge build up on dielectric surfaces by tracking surface charge density ($\sigma$) on boundaries. $\sigma$ at all bounded dielectric locations is updated using the equation below:
\begin{equation}
    \frac{\partial \sigma}{\partial t}=\sum_s q_s \Gamma_s \cdot \hat{n}
\end{equation}
where $q_s$ and $\Gamma_s$ are electrical charge and flux associated with species $s$ and $\hat{n}$ is the surface normal facing away from the gas phase. This surface charge density scaled by surface area to volume ratio is applied to the source term in the Electrostatic Poisson equation (Eq.\ \ref{eq:poisson}) in computational cells next to dielectric boundaries.
\subsection{Boundary conditions}
We use a Maxwellian flux boundary condition for electron density and electron energy at solid boundaries such as electrodes, assuming the electrons are lost at the walls, either through conductors or as trapped charges on dielectric surfaces. The electron flux towards the wall (directed by outward normal $\hat{n}$), is obtained using local number density ($n_e$) and electron temperature ($T_e$), given by:
\begin{equation}
    \Gamma_e \cdot \hat{n}=\frac{n_e}{4} \sqrt{\frac{8 K_B T_e}{\pi m_e}}
    \label{eq:elecflux}
\end{equation}
The electron energy loss flux at solid boundaries is given by:
\begin{equation}
    \Gamma_{E_e} \cdot \hat{n} = 2 k_B T_e \frac{n_e}{4} \sqrt{\frac{8 K_B T_e}{\pi m_e}}
    \label{eq:eenflux}
\end{equation}
Ion flux near solid boundaries is assumed to be mobility dominated. They are assumed to be quenched at conducting walls returning to gas phase as neutrals, while contributing to current through the external circuit. On dielectric walls, they contribute to surface charge build up, while getting quenched to neutral species. The ion flux at solid boundaries is a function of local electric field ($E$) directed towards the surface, ion number density ($n_i$), and ion mobility ($\mu_i$), given by:
\begin{equation}
    \vec{\Gamma}_i \cdot \hat{n}=max\left(\mu_i n_i \vec{E}.\hat{n},0\right)
    \label{eq:ionflux}
\end{equation}
It should be noted that ion flux is 0 if the ion drift flux is directed away from the wall, as given by the conditional expression, in Eq.\ \ref{eq:ionflux}.
Neutral species are also assumed to be quenched at surfaces with the same boundary condition as Eq.\ \ref{eq:elecflux}, except with substitution of neutral number density and gas temperature ($T_g$) instead of electron density and electron temperature ($T_e)$.

These boundary conditions can be easily modified in our solver to accommodate variations such as secondary electron emission (case studied in section \ref{sec:sputter}) in Eq.\ \ref{eq:elecflux} or species interactions with surfaces (e.g., sticking coefficient).
\subsection{Transport properties and chemistry}\label{sec:transchem}
Non-equilibrium plasma chemical mechanisms consists of both electron impact and neutral reactions. The reaction rates of the former is often not in the Arrhenius form. An offline Boltzmann solver (e.g., BOLSIG+ \cite{hagelaar2005solving}) is typically used  to obtain electron-energy-distribution-functions and subsequently closed form polynomial fits for electron impact rate constants as a function of electron temperature or local reduced electric fields ($\bigr\vert\vec{E}\bigr\vert/N$). The neutral reactions can be in both reversible or irreversible Arrhenius form. For the reversible forms, the backward reaction rates are obtained from the computation of the equilibrium constant from reaction Gibb's free energies. The electron transport properties, that include electron mobility and electron energy mobility are also obtained from BOLSIG+ solution as a function of $T_e$ and $\bigr\vert\vec{E}\bigr\vert/N$. The electron and electron energy diffusion coefficients are calculated as a function of corresponding mobilities through the Einstein relation. Ion transport properties are obtained from published atomic and nuclear data tables such as by Ellis et al. \cite{ellis1976transport}.

\section{Numerical Methods}
\label{sec:numerics}
A finite volume formulation on Cartesian grids is used to discretize the plasma governing equations (Eqs.\ \ref{continuity}-\ref{eenergysource}). Each of the equations are cast in the unsteady convection-diffusion-reaction form given by:
\begin{equation}
    \frac{\partial \psi}{\partial t} + \vec{\nabla} \cdot (\psi \vec{V}_{\psi}) = \vec{\nabla} \cdot (D_{\psi} \vec{\nabla} \psi) + S_{\psi}
\end{equation}
where $\psi$ is the conserved variable (e.g., electron density), $\vec{V}_\psi$ is the convection velocity, $D_{\psi}$ is the diffusion coefficient for scalar $\psi$, and $S_{\psi}$ is the source term (e.g., chemical reaction progress rates, space charge, or Joule heating). 
These equations are solved using a segregated procedure with a semi-implicit Backward-Euler time integration scheme.

The convective terms are discretized using a fifth-order Weighted Essentially Non-Oscillatory (WENO) scheme \cite{jiang1996efficient} and is obtained as an explicit source term using conserved variable values from the previous time step. On the other hand, the diffusive flux is obtained using a second order central scheme and is treated implicitly in time. The production term from finite rate chemistry is also treated explicitly in time.

The treatment of non-Cartesian geometry is part of our ongoing efforts. However, we currently include a cell masking formulation where the user can disable solves in parts of the domain while directly specifying flux as well as Dirichlet boundary conditions at interior interfaces through user-defined functions. In this method, a mask field is set as 1 inside valid plasma regions while it is set as 0 in covered regions. The linear solvers are disabled in covered regions with one-sided numerical schemes enabled at covered-valid interfaces.  This approach is accurate for rectangular geometries and can have wide applicability for axisymmetric reactor configurations prevalent in semiconductor fabrication industry. We use this method to simulate the Gaseous Electronics Conference (GEC) reference cell described in section \ref{sec:geccell}.

The time dependent convection-diffusion-reaction system is advanced using a global second-order implicit Runge-Kutta scheme. The mid-point method for solving $\frac{du}{dt}=f(u)$, where $u$ is the set of conserved variables at all finite volume cells, is given by:
\begin{equation}
    u^{n+1}=u^n+f\left(\frac{u^n+u^{n+1}}{2}\right) \delta t
    \label{eq:rk1}
\end{equation}
An equivalent iterative scheme to solve Eq.\ \ref{eq:rk1} can be written using the current iterate $k$ on the right hand side of Eq.\ \ref{eq:rk1} updating the solution at iteration $k+1$, given by:
\begin{equation}
    u^{k+1}=u^n+f\left(\frac{u^n+u^{k}}{2}\right) \delta t
    \label{eq:rk2}
\end{equation}
with $u^0=u^n$, $n$ being the current time-step.
It can be seen from Eq.\ \ref{eq:rk2} that for one iteration, we recover a first-order Explicit-Euler scheme, while with 2 iterations, we recover the second-order Explicit Runge-Kutta scheme. Further iterations beyond 2 (which we refer to as number of time-step correctors $n_t$) continues to improve the accuracy of the method, as will be demonstrated in the upcoming section on time accuracy (section \ref{sec:timeMMS}).
Our solver allows for the use of a fixed time-step specified by the user as well as an adaptive time-step determined by a user specified Courant number ($C$). Electron time-scales tend to be the quickest among all species and therefore used in the calculation of a global adaptive time-step among all cells given by: 
\begin{equation}
    \delta t=\min\left(C_{adv} \frac{\delta x_{min}}{\mu_e |\vec{E|}}, C_{diff} \frac{1}{N_{dim}} \frac{\delta x_{min}^2}{2D_e}, C_{diel} \frac{\epsilon_0}{e \mu_e n_e}\right) 
\end{equation},
where $C_{adv}$, $C_{diff}$, and $C_{diel}$, are user specified Courant numbers for advection, diffusion, and dielectric relaxation time scales. $\delta x_{min}$ is the smallest grid spacing, and $N_{dim}$ is the dimension of the simulation and takes the values from 1,2, or 3 depending on whether the simulation is 1D, 2D or 3D.

\section{Implementation}
\label{sec:implementation}
Our solver, \textit{vidyut3d} \cite{sitaraman2024vidyut3d}, is written in C++ and is developed using the performance portable adaptive Cartesian grid library, AMReX \cite{zhang2019amrex}. It is open-source and can  be accessed at \url{https://github.com/NREL/vidyut3d}. We utilize a non-subcycled approach wherein all adaptive mesh refinement (AMR) levels are advanced with the same time-step. The composite multi-level multigrid solver from within AMReX is used for each of the governing equations that are cast into a Helmholtz equation form. We also interface with HYPRE \cite{falgout2002hypre} for discharge simulations with highly stiff systems, for example when simulating high pressure streamers or discharges with complex boundaries. Stencil operations within the block structured grid utilizes AMReX's C++ lambda-based launch system for performance portable execution on GPUs. Our solver can be used for 1D, 2D, 2D-axisymmetric and 3D simulations which is selected at compile time. 
We include additional polar coordinates derived divergence terms as explicit sources to each of our governing equations in the 2D-axisymmetric mode.
Zero dimensional cases tailored towards complex plasma chemistry can also be performed in the 1D mode using a single or 2 cells (if a linear potential solution with constant electric field is needed).

\begin{figure}
    \centering
    \includegraphics[width=0.6\linewidth]{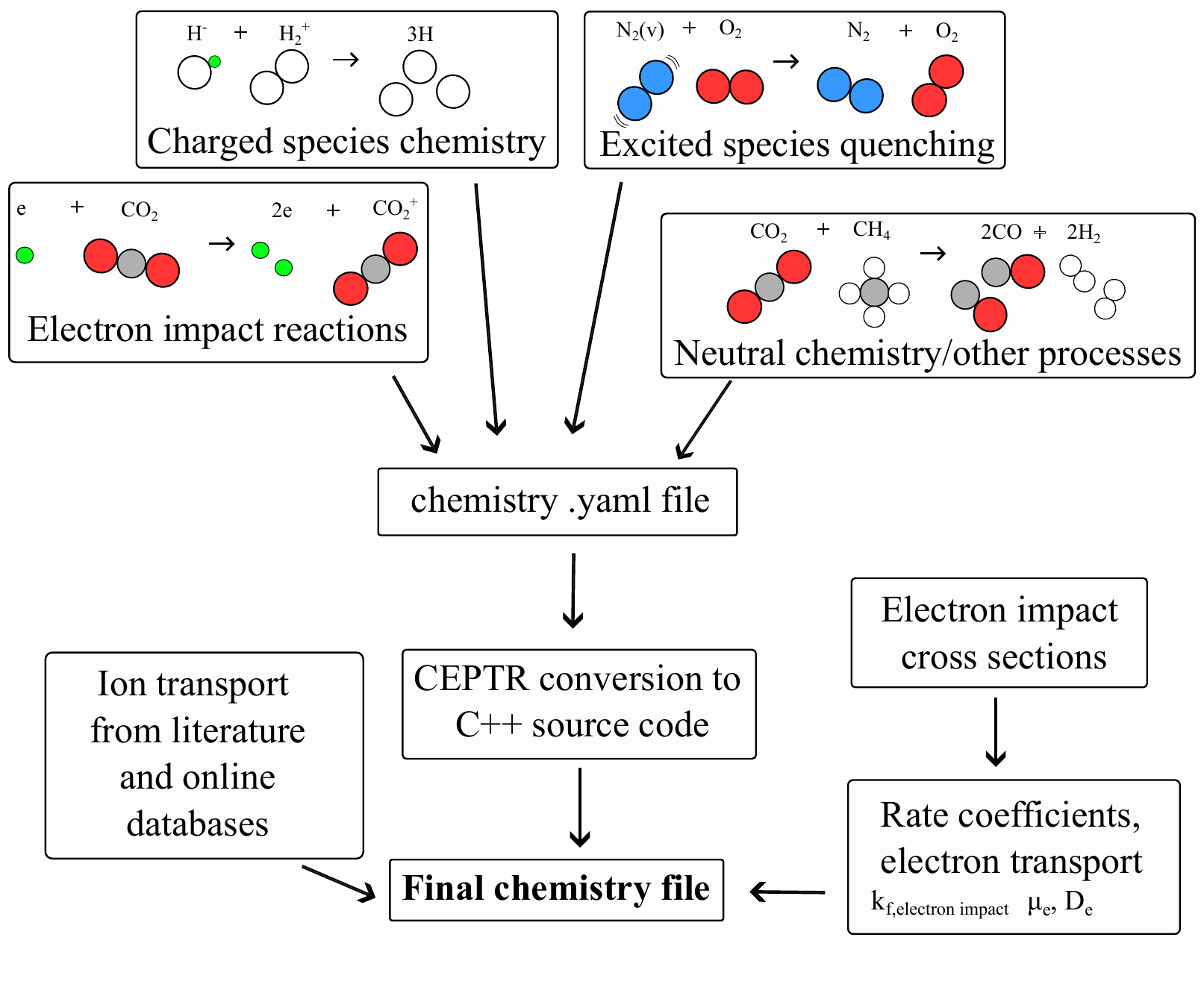}
    \caption{Workflow for incorporating a new plasma chemistry mechanism into \textit{vidyut3d}, beginning with the conceptualization and definition of the plasma and non-plasma pathways that define chemical processes, and ending with generation of CPU/GPU-compatible C++ source code files.}
    \label{fig:plasma_chemistry_workflow}
\end{figure}
We have developed a chemical mechanism parser that uses a similar format as CANTERA~\cite{cantera} yaml files as input. Our python based parser is adapted from \textit{CEPTR}, a lexical converter within the open-source combustion mechanism management library, \textit{PelePhysics} \cite{PeleSoftware,Hassanaly2024}. This parser reads the yaml file and provides a header and a C++ file with transport and production rate functions that can be executed on host (CPU) and device (GPU). A schematic for the workflow involved with incorporating a new plasma mechanism into \textit{vidyut3d} is provided in figure \ref{fig:plasma_chemistry_workflow}. Incorporating new plasma kinetics into \textit{vidyut3d} involves conceptualizing a mechanism that adequately describes the chemistry in scope, often involving electron impact reactions (resulting in e.g,. excitation, dissociation, ionization, etc.), quenching and energy transfers involving excited species (e.g. Vibrational-Vibrational and Vibrational-Translational relaxation), charged species kinetics (e.g. ion-ion and electron-ion recombination, charge transfers), and neutral species chemistry. These reactions along with the necessary thermokinetics data for all species must then be provided in a CANTERA-formatted yaml file. It is important to note that if the mechanism includes reversible reactions involving excited plasma species, special care should be given towards estimating the thermokinetics data for excited species (see Hazenberg et al. \cite{hazenberg2023consistent} for more details on how this can be accomplished). Once the yaml file is complete, our python parser can be used to generate C++ files that can be directly used by \textit{vidyut3d}. As noted in section \ref{sec:transchem}, rate coefficients for electron impact processes are often evaluated using BOLSIG+\cite{hagelaar2005solving}, and parameterized as functions of $T_{e}$ or $\bigr\vert\vec{E}\bigr\vert/N$, and cannot be accurately described in the Arrhenius form. Such reactions, along with transport properties for charged and neutral species must be added to the C++ source code file by the user, while automation of this aspect will be the focus of our future development work.

\section{Code verification and validation}
\label{sec:codeverify}
\subsection{Method of Manufactured solutions (MMS) for spatial accuracy}
\label{sec:mmsspatial}
We utilize the MMS technique to verify the implementation of our solver. We solve the following system in a one dimensional domain  $x \in (0,1)$. This system of equations is a subset of the plasma fluid equations that includes a single species transport equation, drift diffusion approximation and Poisson solve with Dirichlet boundary conditions:
\begin{eqnarray}
\frac{\partial n}{\partial t} + \frac{\partial \Gamma}{\partial x}=\left(\frac{5}{3} x^4 - \frac{x}{6} - 2 \right)\\
\Gamma=\mu n E - D \frac{\partial n}{\partial x}\\
E=-\frac{\partial \phi}{\partial x}\,\,\mu=-1\,\,D=1\\
\frac{\partial^2 \phi}{\partial x^2}=n\\
n(0)=0\,\,n(1)=1\,\,\phi(0)=0\,\,\phi(1)=0
\end{eqnarray}

\begin{figure}
    \centering
    \includegraphics[width=0.9\textwidth]{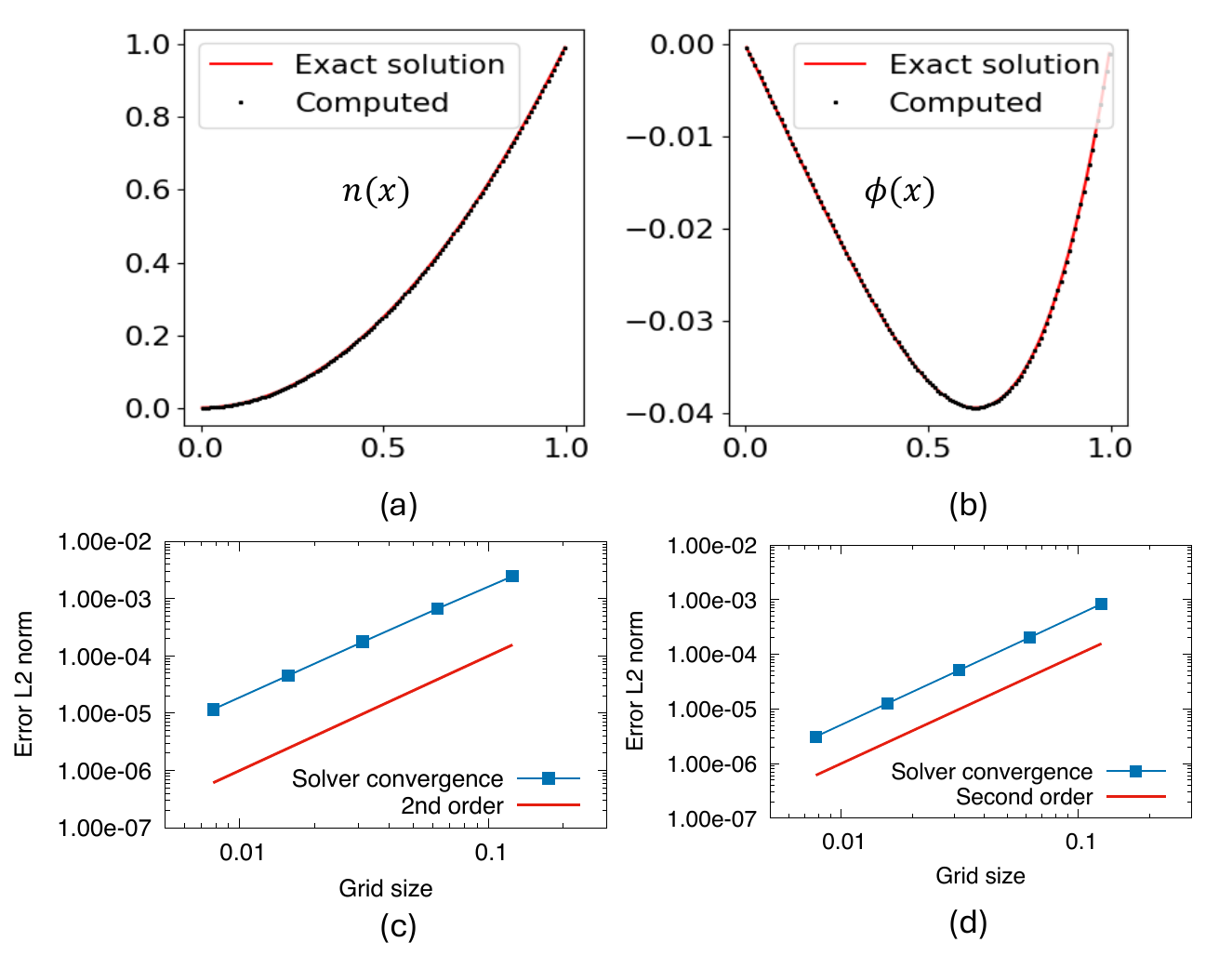}
    \caption{(a) and (b) compare the computed and analytical number density ($n(x)$) and potential ($\phi(x)$) solutions. (c) and (d) show the variation of the error $L_2$ norm with varying grid sizes for $n(x)$ and $\phi(x)$, indicating second-order convergence for the MMS problem described in section \ref{sec:mmsspatial}.}
    \label{fig:mmsconv}
\end{figure}

The exact analytic solution at steady state for this system is as shown below for number density $n$ and potential $\phi$:
\begin{eqnarray}
    n=x^2\\
    \phi=\frac{1}{12}(x^4-x)
\end{eqnarray}
Figure \ref{fig:mmsconv} shows the comparison of computed and analytic solutions on a 128 cell domain. The equations were solved for 100,000 time-steps with a time-step of 1 ms (maximum Courant number $\sim$ 0.5) ensuring a steady-state solution. The solver correctly calculates  the analytic solution and second order accuracy is obtained with varying grid sizes for both $n$ and $\phi$ solutions, respectively. This case therefore verifies our implementation of spatial discretization schemes. It should be noted that even though the advection terms are obtained using a fifth order scheme, the leading order truncation error is from the central diffusion scheme, resulting in a second-order convergence.
\subsection{Method of Manufactured solutions (MMS) for time accuracy}
\label{sec:timeMMS}
This verification case tests the accuracy of our time advance scheme. We consider a one dimension domain in $x \in (0,1)$ with an initial sinusoidal condition for electron/ion density of the form $(1+sin(\pi x))$ in non dimensional units. We set all transport coefficients to 0 in this case so as to isolate convergence in time. A single reaction is modeled in this case with a consumption rate $k_{r}$ equal to 5.0 non dimensional units for all species. The analytic solution to this problem in time is an exponential decay of the initial solution, given by $(1+sin (\pi x)) e^{-k_{r} t}$. 
\begin{figure}
    \centering
    \begin{subfigure}{0.49\textwidth}
    \includegraphics[width=0.99\textwidth]{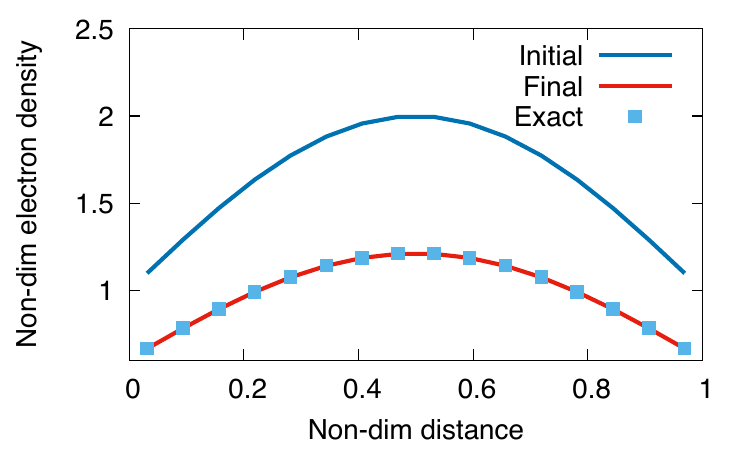}
    \caption{}
    \end{subfigure}
    \begin{subfigure}{0.49\textwidth}
    \includegraphics[width=0.99\textwidth]{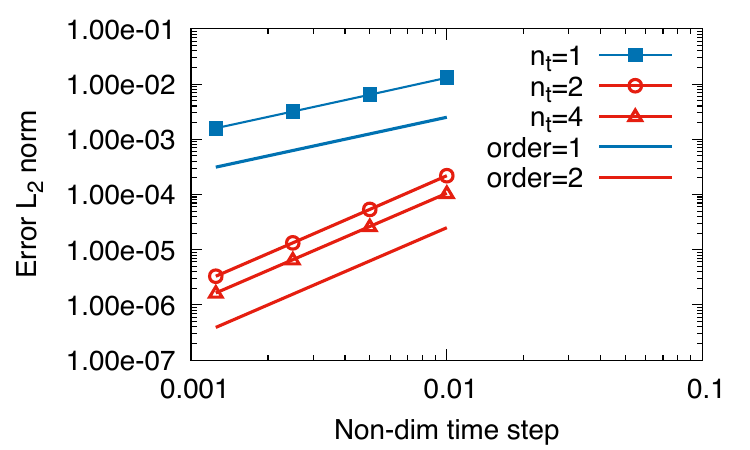}
    \caption{}
    \end{subfigure}
    \caption{(a) Initial and final electron density profile after time $t=0.1$ non-dimensional units along with the exact analytic solution and (b) the $L_2$ norm of electron density error with varying time-step sizes, indicating the expected order of accuracy (first order with $n_t=1$ and second order with $n_t>=2$) with a varying number of time step correctors $n_t$, for the MMS case described in section \ref{sec:timeMMS}.}
    \label{fig:mmstime}
\end{figure}
Figure \ref{fig:mmstime}(a) shows the initial electron density profile and final solution after time $t=0.1$ along with the exact analytic solution, indicating good agreement. Fig.\ \ref{fig:mmstime}(b) shows the error convergence with varying number of time-step correctors ($n_t$). As discussed earlier in section \ref{sec:numerics}, a single iteration ($n_t=1$) shows first order accuracy in time while iterations more than 1 ($n_t>=2$) shows second order accuracy. It can also be seen that lower error in time is obtained with more number of time-step correctors ($n_t=4$) compared to two iterations ($n_t=2$) while maintaining second-order accuracy.
\subsection{Code verification: Capacitive discharge}
\label{sec:ccp}
\begin{figure}
    \centering
    \begin{subfigure}{0.5\textwidth}
    \includegraphics[width=0.99\textwidth]{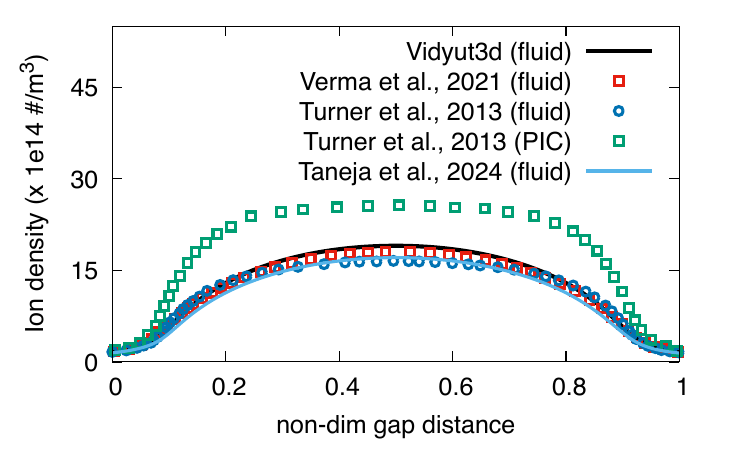}
    \caption{}
    \end{subfigure}
    \begin{subfigure}{0.5\textwidth}
    \includegraphics[width=0.99\textwidth]{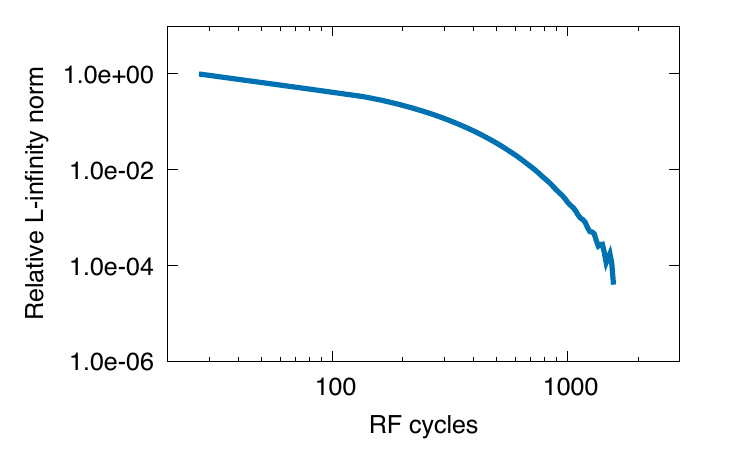}
    \caption{}
    \end{subfigure}
    \begin{subfigure}{0.5\textwidth}
    \includegraphics[width=0.99\textwidth]{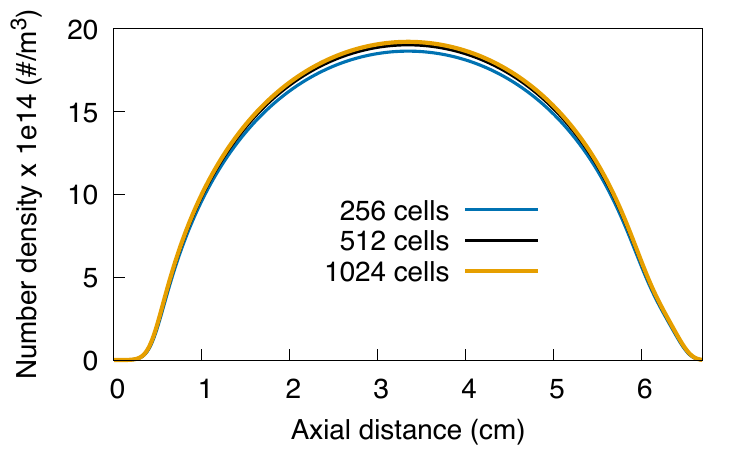}
    \caption{}
    \end{subfigure}
    \caption{(a) Comparison of ion density distribution at steady state (after $\sim$ 1000 RF cycles) with fluid models by Turner et al. \cite{turner2013simulation}, Verma and Venkattraman  \cite{verma2021somafoam}, and Taneja et al. \cite{taneja20241d}, along with the PIC solution by Turner et al. \cite{turner2013simulation}, (b) the relative difference in ion-density $L_{\infty}$ norm between successive time-steps, indicating a reduction in the residual by a factor of 10000 (thus approaching steady state), and (c) ion density distribution from simulations with varying number of cells, indicating grid converged solutions for the capacitive discharge test case described in section \ref{sec:ccp}.}
    \label{fig:ccp}
\end{figure}
In this verification case, we simulated a capacitive discharge in He at 1 Torr and 300 K, similar to benchmark simulations presented by Turner et al. \cite{turner2013simulation}. The helium chemistry used in this case consists of electrons, helium ions and two excited states. The reaction mechanism parameters for electron impact excitation and ionization along with electron/ion transport properties were obtained from recent work by Taneja et al. \cite{taneja20241d}. A sinusoidal voltage waveform at radio frequency (RF) of 13.56 MHz and a voltage of 120 V is applied to one end of a one dimensional domain of distance 6.7 cm while the other end is grounded. A Maxwellian flux is applied at the boundaries for the electron wall loss boundary condition, while mobility dominated loss is applied for ions at the cathodic electrode during every sinusoidal time period. This simulation is performed by building the solver in 1D and run with a uniform mesh of 512 cells. Figure \ref{fig:ccp} shows a comparison of ion density calculated by our solver after $\sim$ 1000 cycles where the peak densities approached a steady state with four orders of magnitude reduction peak ion density residual (Fig.\ \ref{fig:ccp}(b)). Our solution compares well with fluid models documented in the literature by Turner et al. \cite{turner2013simulation}, OpenFOAM based solver, \textit{SOMAFOAM}, by Verma and Venkattraman \cite{verma2021somafoam}, and recent one dimensional solver, \textit{mps1d}, by Taneja et al. \cite{taneja20241d}, and shows a similar discrepancy with PIC models due to functional fits on ion/electron transport properties and electron impact chemistry from offline Boltzmann solves using BOLSIG+. We also performed a grid convergence study to establish the adequacy of our baseline resolution (512 cells). Figure \ref{fig:ccp}(c) indicates almost identical results for ion density profiles with 1024 cells compared to 512 cells after 1000 RF cycles, indicating grid converged results in this case.
%
%
\subsection{Streamer propagation}\label{sec:bagheri}
\begin{figure}
    \centering
    \includegraphics[width=0.99\linewidth]{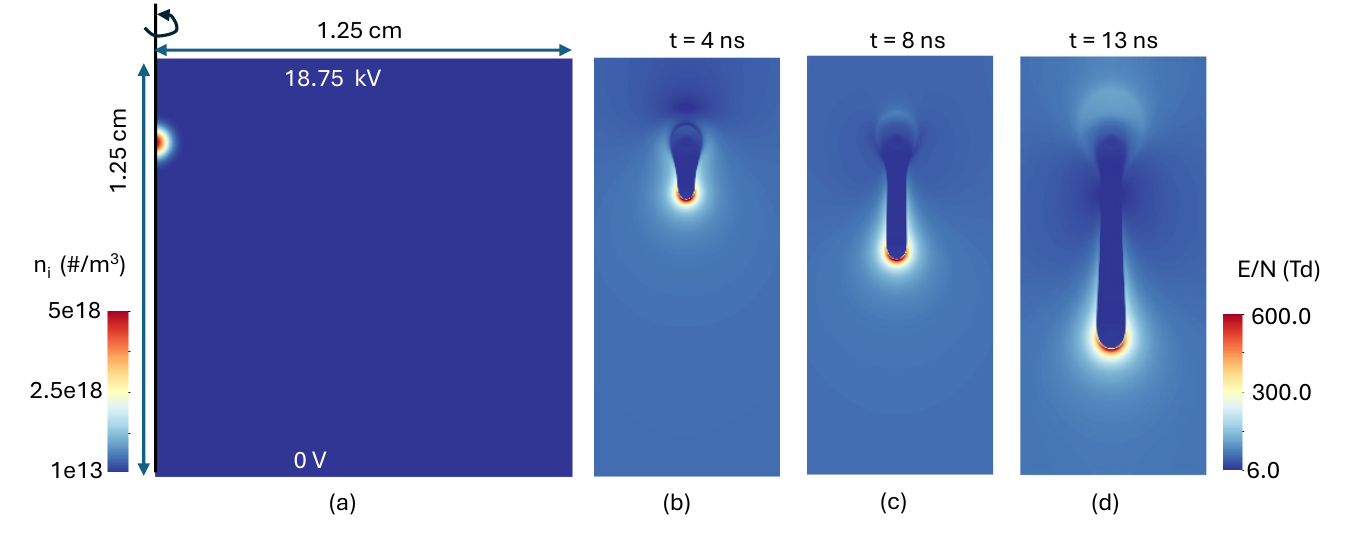}
    \caption{(a) The 1.25 cm x 1.25 cm axisymmetric simulation domain for the streamer test case in section \ref{sec:bagheri} with an initial kernel of ions at a height 1 cm from the bottom boundary, and top and lower boundaries held at voltages of 18.75 and 0 kV, respectively, and (b), (c) and (d) snapshots of the reduced electric fields at 4, 8 and 13 ns, indicating the propagation of a positive streamer.}
    \label{fig:axistreamer}
\end{figure}

We verify our solver with a benchmark positive streamer propagation case presented in the work by Bagheri et al.\ \cite{bagheri2018comparison} where 6 different non-equilibrium streamer codes were compared. This case involves a 2D axisymmetric simulation in a square 1.25 x 1.25 cm domain as shown in Fig.\ \ref{fig:axistreamer}(a). 
\begin{figure}
    \centering
    \includegraphics[width=0.9\linewidth]{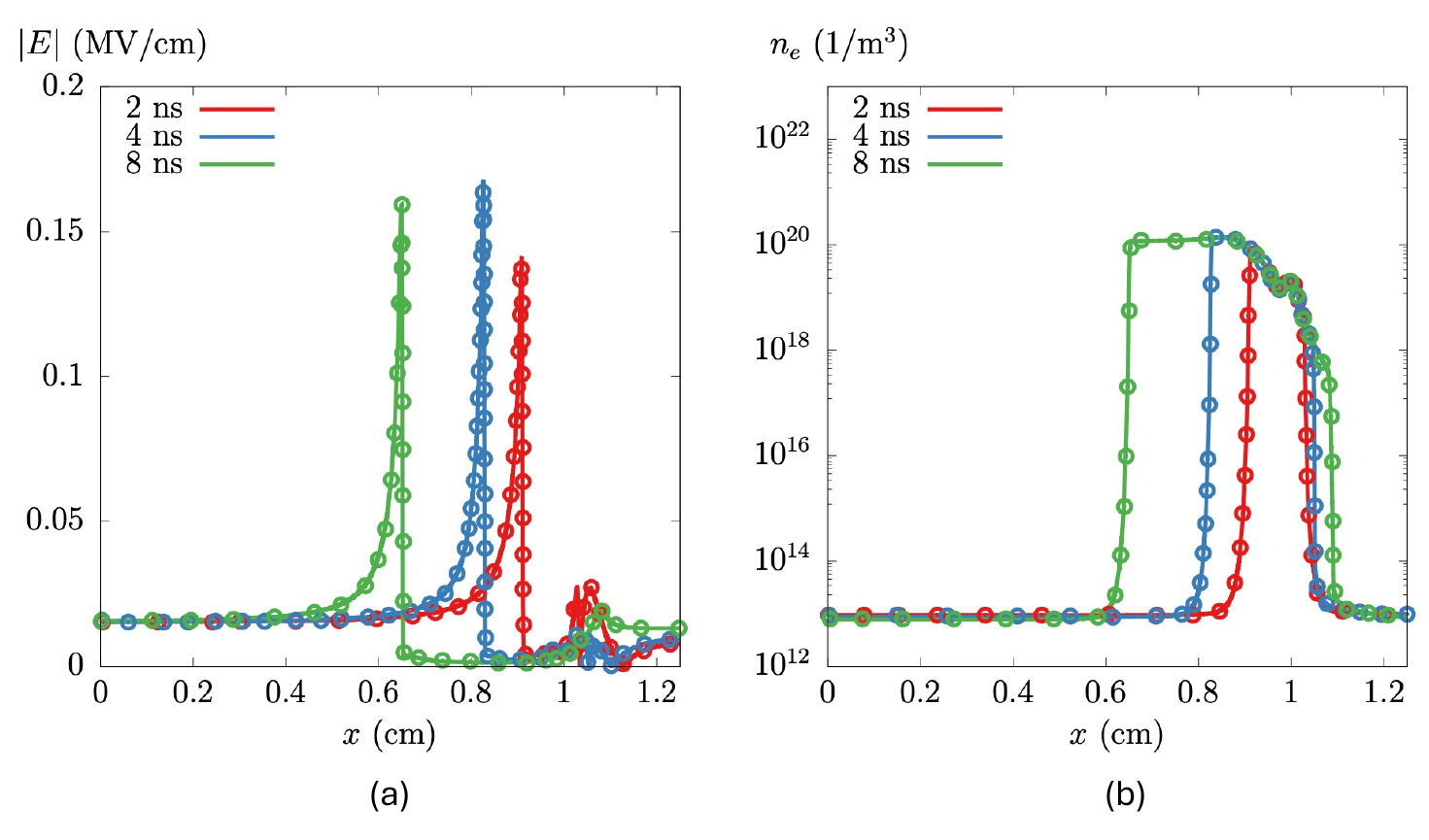}
    \caption{Electric field magnitude (a) and electron density (b) variation along the axis of symmetry at varying times (2, 4, and 8 ns), with solid lines indicating results from our solver and symbols from the work of Bagheri et al. \cite{bagheri2018comparison}, for the streamer test case described in section \ref{sec:bagheri}.}
    \label{fig:axstreamer_profiles}
\end{figure}
%
%
%
\begin{figure}
    \centering
    \includegraphics[width=0.5\linewidth]{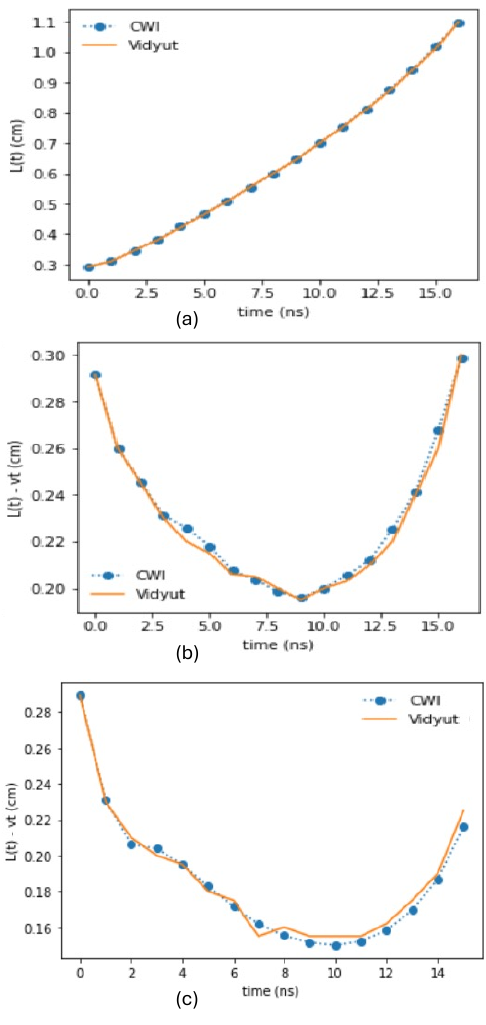}
    \caption{(a) Location of the streamer head, L(t) as a function of time without photoionization physics, (b) $L(t) - vt$ ($v$=0.05 cm/ns) without photoionization and (c) $L(t)-vt$ ($v$=0.06 cm/ns) with photoionization. Results from Bagheri et al.\ \cite{bagheri2018comparison} are denoted as CWI (Centrum Wiskunde and Informatica, Netherlands) in blue, and our results are denoted as \textit{Vidyut} in orange.}
    \label{fig:streamer_head_loc1}
\end{figure}
In this simulation,  an initial seed of ions (Fig.\ \ref{fig:axistreamer}(a)) is placed close to the anodic top boundary that is maintained at a potential of 18.75 kV, while the bottom boundary is at a potential of 0 V. The ion seed in this case provides an initial positive space charge that initiates the streamer propagation in the background gas with uniform electron and ion number density of $10^{13}$\#/m$^{3}$. The electron transport properties and ionization rates are obtained as a function of local electric field magnitude from functional forms presented in Bagheri et al. \cite{bagheri2018comparison}, while the ions are assumed to be stationary over the time scales considered in the test case. The ion evolution in time in this case is only through the electron impact ionization source term. A local field approximation is used in this benchmark case where electron energy equation is not solved, and electron transport properties and electron impact rate coefficients are parameterized as functions of the reduced electric field $\bigr\vert\vec{E}\bigr\vert/N$.  Two levels of AMR were used for this simulation with a base grid of 2048 x 2048 cells, which resulted in a resolution of $\sim$1.5 $\mu$m at the finest level. These additional levels of mesh refinement were applied in regions of the solution that supported large gradients in the electric field and electron number density. A constant time-step of $10^{-13}$ s was used for this simulation. 
Figures \ref{fig:axistreamer}(b),(c) and (d) show snapshots of reduced electric fields indicating streamer propagation towards the cathodic boundary with high electric fields near the space charge dominated streamer head.
Figures \ref{fig:axstreamer_profiles}(a) and (b) show electric field magnitude and electron density profiles along the axis of symmetry, predicted by our solver that compares well against data  from Bagheri et al.\ \cite{bagheri2018comparison}. Similarly, Fig.~\ref{fig:streamer_head_loc1}(a) shows the location of the propagating streamer head as a function of time post-processed from our simulations by finding the location of maximum reduced electric field as a function of time. Figure \ref{fig:streamer_head_loc1}(b) shows a variation of Fig.\ \ref{fig:streamer_head_loc1}(a) where a distance based on an average streamer velocity (0.05 cm/ns) is subtracted from the streamer location, so as to highlight streamer speed variations. There is good agreement between the published results and our solver, thus verifying our numerical implementation of non-equilibrium plasma governing equations.
We also simulated case 3 from Bagheri et al.'s work~\cite{bagheri2018comparison}, where they showed the effect of photoionization on the streamer propagation characteristics by setting a lower background electron and ion density of $10^9$ $m^{-3}$  (which was four orders of magnitude lower than the previous case shown above). The photoionization source term was responsible for creating the extra source of seed electrons downstream of the streamer head that assisted its propagation. 
The three-term Helmholtz-equation model~\cite{bourdon2007efficient} as described in section \ref{sec:goveq} is used here. The streamer head location variation offset by the distance traversed by a constant streamer speed (0.06 cm/ns) has been plotted as a function of time for this case in Fig.~\ref{fig:streamer_head_loc1}(c), similar to Fig.~\ref{fig:streamer_head_loc1}(b). A computational grid with the finest resolution of 1.5 $\mu$m and a time-step of $10^{-13}$ s, same as the case without photoionization was used for this study.  Good agreement was achieved with published results in Bagheri et al.~\cite{bagheri2018comparison}, thus verifying our coupled implementation of photoionization physics and the plasma fluid model. 
\subsection{Gaseous Electronics Conference (GEC) Reference cell}
\label{sec:geccell}
\begin{figure}
    \centering
    \begin{subfigure}{0.55\textwidth}
    \includegraphics[width=0.99\linewidth]{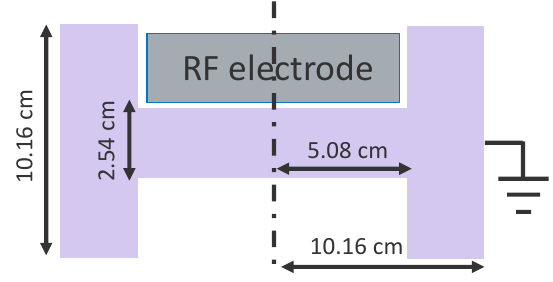}
    \caption{}
    \end{subfigure}
    \begin{subfigure}{0.38\textwidth}
    \includegraphics[width=0.99\linewidth]{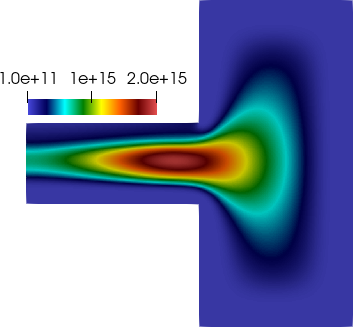}
    \caption{}
    \end{subfigure}
    \begin{subfigure}{0.49\textwidth}
    \includegraphics[width=0.99\linewidth]{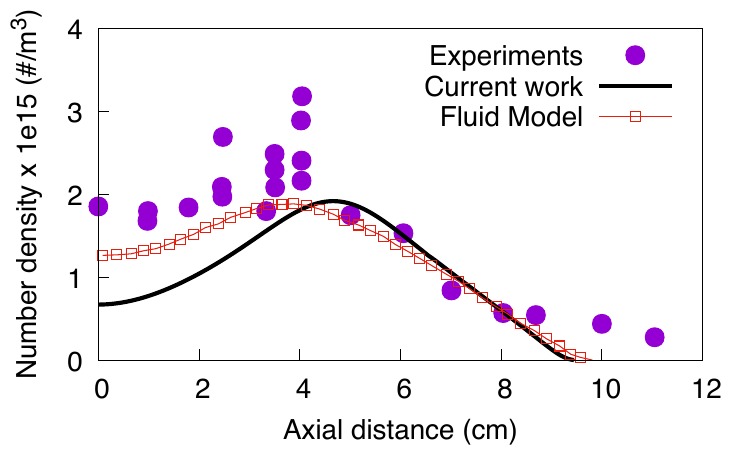}
    \caption{}
    \end{subfigure}
    \begin{subfigure}{0.49\textwidth}
    \includegraphics[width=0.99\linewidth]{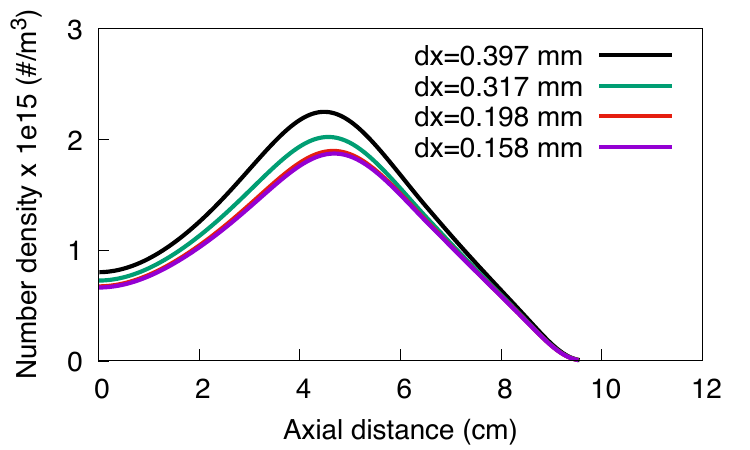}
    \caption{}
    \end{subfigure}
    \caption{(a) Schematic of the GEC reference cell, (b) cycle averaged electron density after 1000 RF cycles, (c) a comparison of the electron density profile along the radial direction at the mid plane against experiments \cite{overzet1993spatial} and the plasma fluid model by Boeuf and Pitchford \cite{boeuf1995two} and (d) the radial electron density profile after 300 RF cycles with 4 different grid resolutions ($256 \times 256$ to $640 \times 640$), indicating a grid converged solution for our base case of $\mathrm{dx=0.198\,mm}$.}
    \label{fig:geccell}
\end{figure}
In this section, we perform a validation study of our solver against the GEC reference cell, which was designed as a baseline for comparison among low pressure plasma experiments for semiconductor applications \cite{brake1999gaseous}. In recent years, the measurements from the GEC cell has been used for validating plasma fluid and particle based models \cite{verma2021somafoam,levko2021vizgrain,li2025fast}. We use the conditions similar to the work by Boeuf and Pitchford \cite{boeuf1995two} for the geometry as shown in Figure \ref{fig:geccell}(a). We use argon gas at a pressure of 100 mTorr and temperature of 300 K in these 2D axisymmetric  simulations. The argon plasma chemistry includes two ionic species ($\mathrm{Ar^+}$ and $\mathrm{Ar_2^+}$) and two metastable species ($Ar^*$ and $Ar_2^*$); the reaction mechanism including electron impact reaction rates are shown in table \ref{Ar_rxns} in \ref{paramsArH2}. The voltage at the powered electrode is sinusoidal in time with a peak value of 100 V and a frequency of 13.56 MHz while other boundaries except the axis are grounded. We use a $\mathrm{10.16 \times 10.16}$ cm cartesian domain resolved by a $\mathrm{512 \times 512}$ grid with the cell-masking procedure outlined in section \ref{sec:implementation} to create the axisymmetric geometry shown in Fig.\ \ref{fig:geccell}(b). This simulation was performed on a uniform grid without AMR with a resolution of 0.198 mm with the base $\mathrm{512 \times 512}$ grid. A Maxwellian flux for both the electron density and electron energy equations are imposed at all boundaries except the axis of symmetry using conditions described by Eqs.\ \ref{eq:elecflux} and \ref{eq:eenflux}, respectively. Mobility dominated fluxes for ions given by Eq.\ \ref{eq:ionflux} are imposed at all boundaries. The solver was run for 1000 RF cycles with a fixed time step of $\mathrm{5 \times 10^{-12}}$ seconds to achieve steady ion densities similar to helium RF case shown earlier (Fig.\ \ref{fig:ccp}(b)). Figure \ref{fig:geccell}(b) shows cycle averaged electron density distribution indicating a near symmetric profile with a central maximum that is offset towards the outer region of the cell. This profile is very similar to other documented computational studies in literature \cite{verma2021somafoam, levko2021vizgrain, chen2024electrical}. Figure \ref{fig:geccell}(c) shows the mid-plane radial distribution of electron densities  calculated from our simulation along with experimental measurements by Overzet and Hopkins \cite{overzet1993spatial} and the fluid model by Boeuf and Pitchford \cite{boeuf1995two}. Reasonable agreement with experiments is achieved here with our simulations predicting peak electron density values of $\sim$ $\mathrm{2 \times10^{15} \#/m^3}$. Our fluid model shows deviations from the results of Boeuf and Pitchford \cite{boeuf1995two} due to the inclusion of electron temperature and reduced electric field dependent transport properties, non-Arrhenius reaction rates, and a larger argon plasma kinetic mechanism. Figure \ref{fig:geccell}(d) shows a grid convergence study with varying grid sizes including our base grid size (0.198 mm) indicating a converged solution.
\section{Demonstration simulations}
\label{sec:demo}
\subsection{Streamer discharge in $\ce{Ar-H2}$ mixtures: 3D simulations}
\label{sec:3dstreamer}
\begin{figure}
    \centering
    \includegraphics[width=0.99\linewidth]{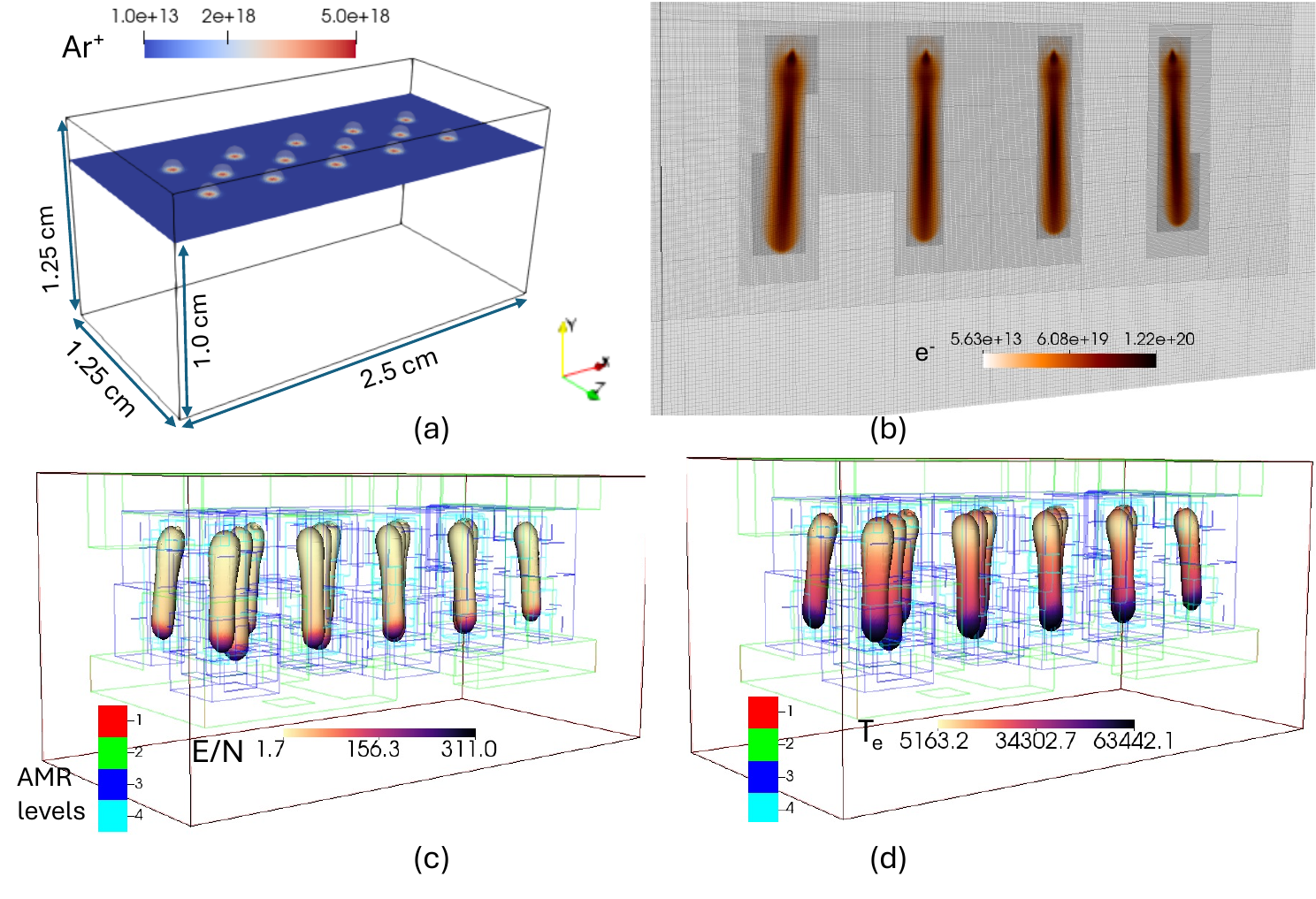}
    \caption{(a) Schematic of the domain with 14 seed charge regions for the \ce{Ar}-\ce{H2} streamer discharge simulation presented in section \ref{sec:3dstreamer}, (b) electron density in units of $\mathrm{\#/m^3}$ and the adaptive grid on a mid plane slice normal to the z direction, (c) the reduced electric field in units of Td and (d) the electron temperature in units of K, at time $t=5\,\mathrm{ns}$ on an electron density isosurface at $\mathrm{10^{19} \#/m^3}$.}
    \label{fig:ArStreamer1}
\end{figure}
To demonstrate adaptive meshing capabilities of our solver, a three-dimensional simulation 
comprised of 14 interacting streamers is performed. The domain is a 2.5 x 1.25 x 1.25 cm$^{3}$ box 
with top and bottom boundaries ($y$-direction) held at fixed potentials. The background gas in this 
simulation is a mixture of 95\% argon and 5\% hydrogen (by volume) at 
atmospheric pressure (101325.0 Pa) and room temperature (300 K). Such mixtures are actively being studied for plasma assisted metal oxide reduction and refining applications \cite{satritama2024hydrogen}. The plasma kinetics mechanism 
used here is given in \ref{paramsArH2} and includes 15 species and 33 reactions with 6 ions ($\ce{Ar+,Ar_2^+,
H+,H_2^+,H_3^+,ArH^+}$) and 
6 neutrals including metastable ($\ce{Ar^*,Ar_2^*}$), vibrationally excited ($\ce{H_2(v1-v3)}$), and dissociated states ($\ce{H}$), alongside background $\ce{Ar}$ and $\ce{H2}$.
To study the impact of interacting streamers, 14 spherical kernels of $\ce{Ar+}$ ions 
are initialized next to each other near the anode as shown in Fig.\ \ref{fig:ArStreamer1}.
The potential at the top boundary is at 18.75 kV while the bottom boundary is kept at 0 V, same as the streamer verification studied earlier (section \ref{sec:bagheri}). 
The 14 kernels are centered at $y=10$ mm with 3 rows separated by a distance of $\delta z=2$ mm 
with $x$ locations spaced out equally at distance of 4.25 mm.
The Gaussian profile from the earlier air streamer verification case by 
Bagheri et al.\ \cite{bagheri2018comparison} from section \ref{sec:bagheri} is used here, with a peak ion density of $\mathrm{5 \times 10^{18} \#/m^3}$ and Gaussian radius of 0.4 mm. 
The domain is discretized using a base level grid comprised of $\mathrm{256\times128\times128}$, 
yielding a grid spacing of $\delta x_{0} \approx 0.1$ mm. 
To adequately resolve the streamers, a total of 4 AMR levels were used  
with refinement added in regions with large reduced electric fields ( $>$ 300 Td)  
and electron number densities ( $> \mathrm{2 \times 10^{19} \#/m^3}$) thresholds, yielding a minimum grid spacing $\delta x_{\text{min}} \approx 12$ $\mu$m. 
The adaptive mesh levels used to refine the propagating streamers at $t=5$ ns are 
shown in Fig. \ref{fig:ArStreamer1}(b)-(d). Fig.\ \ref{fig:ArStreamer1}(b) shows electron density on a mid-plane slice along z direction indicating the propagation of the 4 centrally located streamers that are resolved by additional AMR levels. Figures \ref{fig:ArStreamer1}(c) and (d) show isosurfaces of electron density at $\mathrm{10^{19} \#/m^3}$ colored by reduced electric field and electron temperature, respectively, with the various AMR level blocks resolving each streamer. Peak reduced electric fields $\sim$ 300 Td and electron temperatures $\sim$ 5.5 eV are observed at the streamer heads arising space charge regions.  A total of 337 million cells were present at the end of 9 ns with 61\% of the domain covered by the finest level (level=4), 30\% by level 3, 6.5\% by level 2, and only about 2.5\% by the base level (level=1). This simulation took about 3 hours of wall-clock time with 200 NVIDIA-H100 GPUs to run 9 ns with a fixed time-step of $10^{-12}$ seconds. 
\begin{figure}
    \centering
    \includegraphics[width=0.99\linewidth]{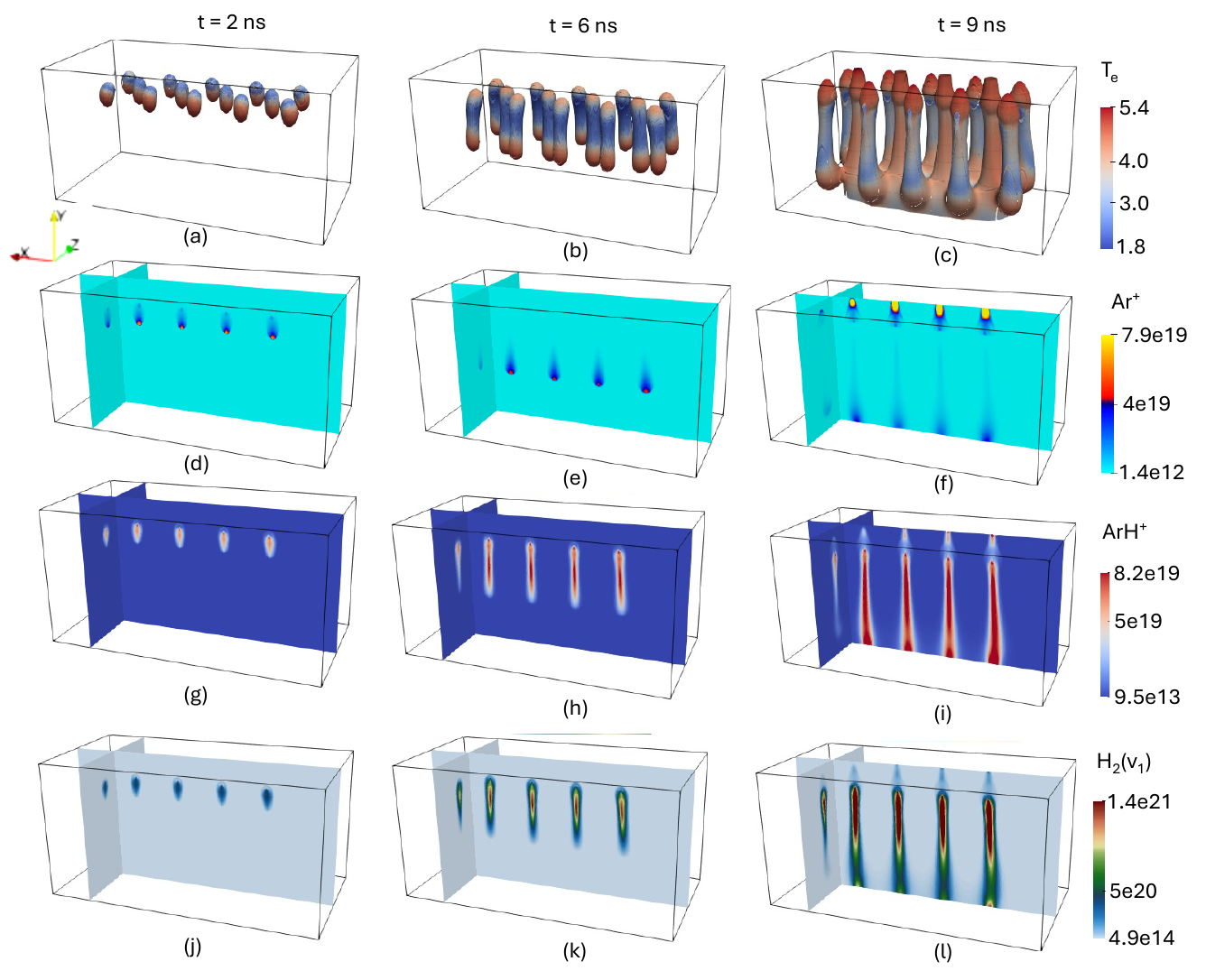}
    \caption{Temporal snapshots of streamers in the \ce{Ar}-\ce{H2} environment at 2,6 and 9 ns from the simulation case described in section \ref{sec:3dstreamer}: (a)-(c) show the electron density isosurface at $\mathrm{1 \times 10^{18} \#/m^3}$  colored by electron temperature, (d)-(f) show \ce{Ar^+}, (g)-(i) show \ce{ArH^+}, and (j)-(l) show \ce{H2(v_1)} densities in units of $\mathrm{\#/m^3}$, respectively.}
    \label{fig:ArStreamer2}
\end{figure}
The streamer propagation transients are shown in Fig.\ \ref{fig:ArStreamer2} at three different time instances (2, 6, and 9 ns), indicating non-uniform ionization wave speeds among the 14 streamers due to inhomogeneous interacting electric fields. Figures \ref{fig:ArStreamer2}(a)-(c) show electron density isosurfaces at $\mathrm{1 \times 10^{18} \#/m^3}$ colored by electron temperature. The streamers in the middle row tend to propagate faster and reach the grounded bottom boundary while the streamers on the side rows merge into the near cathode zone. The anode directed streamers are also formed towards the top boundary due to the higher potential from the space charge. Figures \ref{fig:ArStreamer2}(d)-(f),  \ref{fig:ArStreamer2}(g)-(i), and \ref{fig:ArStreamer2}(j)-(l), show \ce{Ar^+}, \ce{ArH^+} and \ce{H2(v_1)} number densities on two slices of the domain (z and x directions), respectively. \ce{ArH^+} is observed to be the dominant ion. \ce{Ar^+} is seen to be consumed in the bulk of the streamer while being present only at the streamer heads where ionization pathways are dominant. Significant concentration of \ce{H2} vibrational states are observed in the paths of the streamers indicating dominant electron impact excitation pathways.
%
%
%

%
\subsection{RF three-electrode reactor: 3D simulations}
\label{sec:sputter}
\begin{figure}
    \centering
    \includegraphics[width=0.99\linewidth]{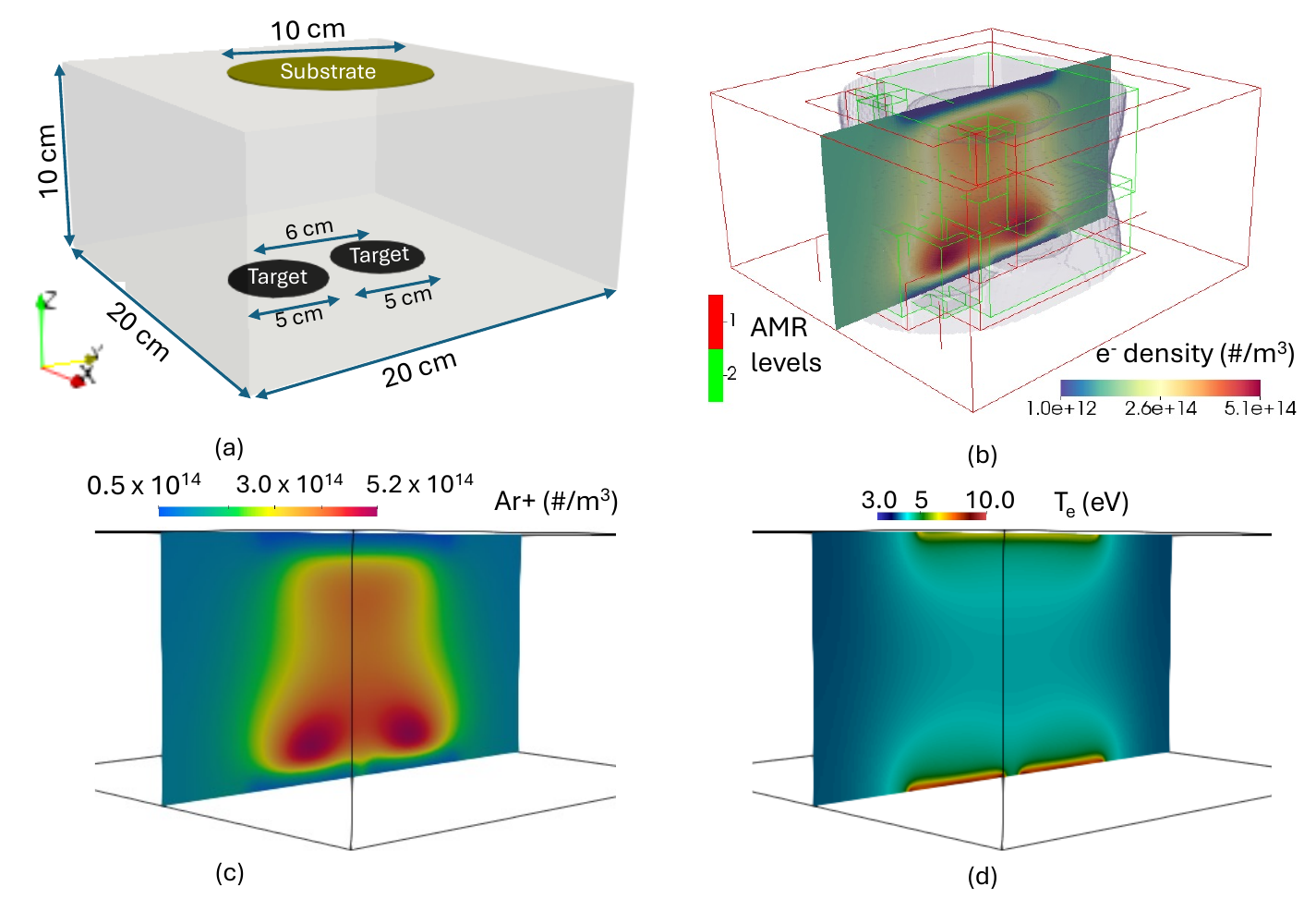}
    \caption{(a) Schematic of the RF three-electrode reactor described in section \ref{sec:sputter}, (b) electron density on a mid plane along the x direction along with level 1 and level 2 AMR blocks, and  (c) and (d) cycle averaged number density of \ce{Ar^+} and electron temperature distribution, on a mid-plane slice along the x direction.}
    \label{fig:sputter}
\end{figure}
We describe simulations of a low pressure three-electrode reactor in this section that is used for thin film deposition, to demonstrate the capability of our solver for low pressure applications. Figure \ref{fig:sputter}(a) shows the geometry of the system studied in this section, which is similar to the reactor experimentally studied by Febba et al.\ \cite{febba2023autonomous}. There are two electrodes at the bottom boundary that represent sputtered targets in the reactor powered with an RF voltage waveform with 60 V amplitude and 13.56 MHz frequency. The top boundary consists of a larger electrode that is grounded representing the substrate. A homogeneous Neumann condition is applied at the rest of the boundaries for potential, species, and electron energy equations, while the surface flux condition for electrons, ions, neutrals, and electron energy is applied at the electrodes (Eqs.\ \ref{eq:elecflux}-\ref{eq:ionflux}).  We perform 3D simulations with pure $\ce{Ar}$ as the background gas at 0.1 torr pressure and 300 K temperature without the presence of metal ions in the gas phase as an effort towards an ongoing experimental validation. We use 2 adaptive mesh refinement levels in these simulations as indicated by the blocks in Fig.\ \ref{fig:sputter}(b). Secondary electron emission is included in these simulations with secondary electron emission coefficient $\gamma=0.1$ for all electrode boundaries, which is a typical value for metals \cite{lieberman1994principles}. These simulations were run for approximately 1000 RF cycles until a steady-state in ion densities were achieved. The number of cells among both AMR levels stabilized around 3.4 million cells after about 100 cycles. Adaptive time-stepping was used in this simulation with Courant numbers, $C_{adv}=0.3$, $C_{diff}=10.0$, and $C_{diel}=10.0$. Time steps were observed to remain between $\mathrm{3 \times 10^{-11}}$ and $\mathrm{7 \times 10^{-11}}$ seconds with the aforementioned Courant numbers. A total time of $\sim$ 192 hours were required to simulate 1000 RF cycles with 256 CPU cores. Fig.\ \ref{fig:sputter}(b) shows an electron density snapshot at the positive half cycle peak and (c) shows cycle-averaged \ce{Ar^+} density on a mid-plane slice, respectively.  Higher plasma densities are observed close to the powered electrodes at the bottom boundary due to the asymmetric potential distribution and focusing in the region between the targets in this geometry. Figure \ref{fig:sputter}(b) also includes a volumetric region (shaded in purple) that indicates the zone where the $\ce{Ar^*}$ densities are between $\mathrm{5 \times 10^{14}}$ and $\mathrm{5 \times 10^{15} \#/m^3}$ where the plasma is active and mesh adaptivity is used. Figure \ref{fig:sputter}(d) shows time-averaged electron temperature indicating the formation of plasma sheaths at the electrodes with temperatures of approximately 10 eV.
\section{Computational performance}
\label{sec:perf}
We used three different machines with varying architectures to assess computational performance of our solver. The configurations of these machines are as shown in table \ref{tab:machines}.
\begin{table}[ht]
    \centering
    \begin{tabular}{|p{0.1\linewidth}|p{0.27\linewidth}|p{0.2\linewidth}|p{0.3\linewidth}|}
        \hline
        \hline
         Machine & CPU hardware & GPU hardware & Compilers \\
         \hline
         \hline
         M1 & 128 $\times$ AMD EPYC-Genoa cores & 4 $\times$ NVIDIA H100 GPUs& gcc 12.2.1, cray-mpich 8.1, cuda 12.3 \\ 
         \hline
         M2 & 60 $\times$ AMD EPYC-Rome cores & 2 $\times$ NVIDIA A100 GPUs & gcc 12.3, mpich 4.2.1, cuda 12.4\\
         \hline
         M3 & 48 $\times$ AMD EPYC-74F3 cores & 8 $\times$ AMD MI250X GPUs & gcc 13.2, openmpi 3, rocm 6.2.4\\
         \hline
    \end{tabular}
    \caption{Compute node specifications for performance tests.}
    \label{tab:machines}
\end{table}
We compiled our solver using the available GNU C compilers along with \textit{mpich/openmpi} \cite{gropp1996user,gabriel2004open} implementations on the respective machines. Machines M1 and M2 have NVIDIA GPUs while M3 has AMD GPUs, thus requiring \textit{hipcc} for compilation as opposed to \textit{nvcc} on M1 and M2. The ability to build and run on AMD as well as NVIDIA GPUs demonstrates the performance portability of our solver.

We performed three different tests to assess computational performance of our solver on CPUs and GPUs. We used the 3D streamer case from section \ref{sec:3dstreamer} with 14 interacting streamers and performed 10 time steps and used the average wall clock time per time step as the metric for comparison studies. 
\begin{figure}
    \centering
    \includegraphics[width=0.99\linewidth]{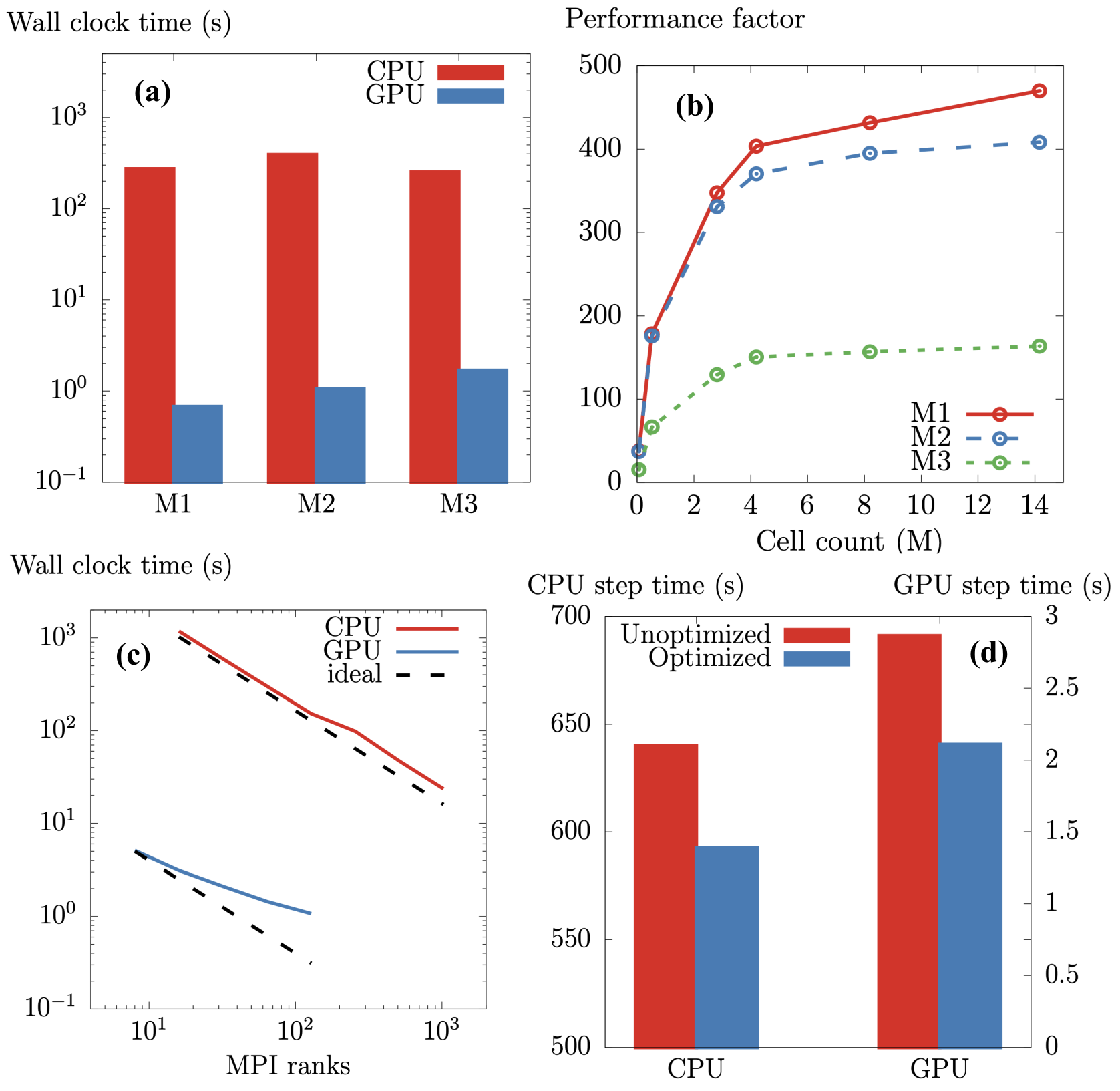}    \caption{(a) Wall clock time per time-step for serial runs on CPUs and GPUs for the 3D streamer case (section \ref{sec:3dstreamer}) with 4 million cells, (b) performance improvement factor for 1 GPU vs 1 CPU core with varying problem sizes from 65k to 15 M cells, (c) strong-scaling of a simulation with 268 million cells over multiple distributed processors with total wall clock time for 10 time-steps, and (d) the performance improvement factor on CPU and GPU with 32 MPI ranks and 268 million cells  before and after code optimizations that include GPU-aware-MPI, multi-component species solves, and memory access improvements. }
    \label{fig:gpucpu}
\end{figure}

Fig.\ \ref{fig:gpucpu}(a) shows wall clock time per timestep for a serial run (single MPI rank) of our solver on a uniform $\mathrm{256 \times 128^2 \sim 4\,million}$ cell grid. The GPU execution times are significantly faster compared to CPU, yielding speedups of two orders of magnitude or more. Machine M1, which is the USA National Renewable Energy Laboratory's (NREL) supercomputer \textit{kestrel} \cite{kestrel-ref}, is observed to be the quickest among the three machines studied in this work. The NVIDIA H100 GPU in this case is observed to be faster than the A100 and the MI250 GPUs by a factor 1.5 and 2, respectively. Overall, a single GPU is about 150-400 times faster than a single CPU core on the three machines studied in this work, for a problem size of 4 million cells.

The speed-up from GPUs compared to CPUs strongly depends on the ratio between execution effort vs memory management and traffic. Higher arithmetic intensity is favored on GPUs and larger problem sizes per GPU will allow execution time to dominate over memory allocation, transfer, and other performance loss mechanisms. Fig.\ \ref{fig:gpucpu}(b) shows the performance improvement factor as a function of problem size from using a single GPU over a CPU for the three machines studied in this work.  Performance improvement factor is higher for larger problem sizes where the execution times dominate. Approximately 500X improvement is observed for large problem sizes $\sim$ 14 million cells while $\sim$ 30-40X is seen with $\sim$ 65,000 cells. Further optimizations regarding memory management and transfer are necessary to improve GPU performance at lower counts, which is part of our ongoing efforts. 

Strong scaling performance across distributed processors using message-passing-interface (MPI) is shown in Fig.\ \ref{fig:gpucpu}(c) for machine M1, for the 3D argon streamer problem on a uniform $\mathrm{1024 \times 512^2=268\,million}$ grid. The average wall clock time is shown on the y axis for the execution of 10 time steps. Favorable strong scaling is observed on the CPU through 1024 MPI ranks. The GPU scaling, although about two orders of magnitude faster than the CPU, shows sub-optimal strong scaling mainly from performance overheads of CPU-GPU memory transfer as well as inter GPU message exchanges, with performance degradation beyond 40 GPUs. 

Finally, recent efforts to increase the overall efficiency of the code, which include improvements to memory access, use of GPU-aware-MPI, as well as the simultaneous solution of multiple species with each multi-level multigrid solve, are demonstrated in Fig.\ \ref{fig:gpucpu}(d). This figure shows the change in the average wall clock time per time step on \textit{Kestrel} for the 3D argon-hydrogen streamer problem (section \ref{sec:3dstreamer}) on a uniform $\mathrm{1024 \times 512^2=268\,million}$ grid, using 32 MPI ranks and both CPU and GPU parallelization strategies. The CPU case shows a 7.5\% increase in computational efficiency, while the GPU case shows a 26\% increase in computational efficiency. One reason for the greater increase in GPU performance as compared with CPU performance is the fact that simultaneously solving multiple scalar solution components with each multigrid solve results in more GPU work, and thus better utilization of the hardware. Lastly, the speedups yielded by these code optimizations is highly problem-dependent, whereby problems with larger and more complex chemistry mechanisms will demonstrate greater speedup.

\section{Conclusions and future work}
In this work, we presented the numerical methodology, solver verification, and computational performance of a GPU compatible fluid solver for non-equilibrium plasmas. Equations for electron and heavy species conservation, along with self-consistent Poisson and electron energy are solved on adaptive Cartesian grids. Our implementation is open-source and can be accessed at \url{https://github.com/hsitaram/vidyut3d}. Our code is verified against five cases: two method of manufactured solutions (MMS) problems, 1 Torr He capacitive discharge, an atmospheric pressure streamer problem with and without photoionization, and the GEC RF cell. The accuracy of our numerical methods, verified using MMS, showed formal second order accuracy in both space and time. Good agreement of our simulation predictions with published results for plasma density and electric fields were obtained for the He capacitive discharge, streamer propagation, and the GEC RF cell. We demonstrated the use of our code on two 3D simulations: atmospheric pressure streamers in \ce{Ar}-\ce{H2} mixtures as well as a three electrode RF thin film deposition reactor.  Computational performance studies indicated $\sim$ 150-400X   performance gain from using a single GPU versus a single CPU core for production scale simulations ($\sim$ 4 million cells). Strong-scaling parallel performance over multiple CPUs/GPUs for a 268 million cell 3D streamer problem indicates favorable scaling up to 2048 CPU cores while up to 100 MPI ranks when using GPUs. 

Our ongoing efforts are along inclusion of geometries using a volume penalization method, current/voltage control using external circuits, experimental validation, and improving GPU performance via profiling and optimizations.

\section*{Acknowledgments}
 This work was authored in part by the National Renewable Energy Laboratory, operated by Alliance for Sustainable Energy, LLC, for the U.S. Department of Energy (DOE) under Contract No. DE-AC36-08GO28308.  
 This material is based upon work supported by the U.S. Department of Energy (DOE), Office of Science, Office of Basic Energy Sciences (BES), Materials Sciences and Engineering Division under Award DE-SC0024724 “Fundamental Studies of Hydrogen Arc Plasmas for High-efficiency and Carbon-free Steelmaking” and U.S. Department of Energy's Laboratory Directed Research and Development (LDRD). The views expressed in the article do not necessarily represent the views of the DOE or the U.S. Government. The U.S. Government retains and the publisher, by accepting the article for publication, acknowledges that the U.S. Government retains a nonexclusive, paid-up, irrevocable, worldwide license to publish or reproduce the published form of this work, or allow others to do so, for U.S. Government purposes.
\newpage

\bibliography{refs}
\bibliographystyle{elsarticle-num}

\appendix
\section{Plasma chemistry: \ce{Ar} and \ce{H2} mixtures}
\label{paramsArH2}
The finite rate chemical mechanism for \ce{Ar} and \ce{H2} plasma chemistry is presented here in the form of tables with associated reaction rate expressions and energetics. We also present electron and ion transport properties in Table \ref{ArH2_mob}, that are used in simulations shown in this work.
\begin{longtable}{|p{0.05\linewidth}|p{0.22\linewidth}|p{0.47\linewidth}|p{0.2\linewidth}|}
\hline
\hline
No.&Reaction& Rate Coefficient (m, \#)& Energy (eV) \\
\hline
\hline
R1&\ce{e + Ar -> Ar^* + e}&$\exp(-37.44+1.407\,log(\epsilon_e)+50.34/\epsilon_e
                                         -348.6/\epsilon_e^2-230.5/\epsilon_e^3)$& 11.6\\
\hline
R2&\ce{e + Ar -> Ar^{+} + 2e}&$\exp(14.37-8.641\,log(\epsilon_e)-622.9/\epsilon_e+
                       5341.0/\epsilon_e^2
                       - 0.1862e5/\epsilon_e^3)$& 15.759\\
\hline
R3&\ce{e + Ar^* -> Ar^+ + 2e}&$\exp(-5.5e28/T_e^6
                                         +5.08e24/T_e^5
                                         -1.85e20/T_e^4
                                         +3.38e15/T_e^3
                                         -3.34e10/T_e^2
                                         +1.39e5/T_e-29.7)$ & 4.159 \\ \hline
R4&\ce{Ar^* + Ar^* -> e + Ar + Ar^+}&$5.0 \times 10^{-16}$& \\ \hline
R5&\ce{Ar^* + e -> Ar + e} & $\exp(-4.82e4/T_e-32.46)$ & \\
\hline
R6&\ce{Ar_2^* + e -> Ar_2^+ + e}&$1.29 \times 10^{-16}\,T_e^{0.7} \exp(-0.4245e5/T_e)$& 4.2\\ \hline
R7&\ce{Ar_2^* + e -> Ar + Ar + e}& $10^{-13}$ & \\ \hline
R8&\ce{Ar^* + 2Ar -> Ar_2^* + Ar}&$1.14 \times 10^{-44}$& \\ \hline
R9&\ce{Ar^+ + 2Ar -> Ar_2^+ + Ar}&$2.5 \times 10^{-43}$&  \\ \hline
R10&\ce{Ar_2^* -> 2Ar}&$6 \times 10^7$&  \\ \hline
R11&\ce{2 Ar_2^* -> Ar_2^+ 2Ar + e}&$5 \times 10^{-16}$&  \\ \hline
R12&\ce{Ar^+ + e -> Ar^*}&$43 \times 10^{-18} T_e^{-0.5}$&  \\ \hline
R13&\ce{Ar^+ + 2e -> Ar^* + e}&$9.75 \times 10^{-21} T_e^{-4.5}$& \\ \hline
R14&\ce{Ar_2^+ + e -> Ar^* + Ar}&$25.9 \times 10^{-12} T_e^{-0.66}$&  \\ \hline
\caption{Reactions, rate coefficients (final reaction rates in $\mathrm{\#/m^3/s}$) and electron energy loss, for \ce{Ar} plasma chemistry. Rate coefficients for R1 and R2 use mean energy, $\epsilon_e=\frac{3}{2} T_e$, where $T_e$ is the Electron temperature in eV. R3-R14 rate coefficients use $T_e$ in units of K. These rates are obtained from Sharma et al. \cite{sharma2016effect} and the work by Levko and Raja \cite{levko2021self}.} 
\label{Ar_rxns}
\end{longtable}

\begin{longtable}{|p{0.05\linewidth}|p{0.22\linewidth}|p{0.5\linewidth}|p{0.2\linewidth}|}
\hline
\hline
No.&Reaction& Rate Coefficient (m, \#)& Energy (eV) \\
\hline
\hline
R15&\ce{e + H2 -> 2H + e}&$8.4 \times 10^{-14} T_e^{-0.45} \exp(-11.18/T_e)$& 14.68\\
\hline
R16&\ce{e + H -> H^{+} + 2e}&$1.1 \times 10^{-14} T_e^{0.29} \exp(-15.28/T_e)$& 13.6\\
\hline
R17&\ce{e + H2 -> H_2^+ + 2e}& $2.3 \times 10^{-14} T_e^{0.19} \exp(-17.87/T_e)$ & 15.4 \\ \hline
R18&\ce{H2 + e -> H+ + H + 2e}&$9.4 \times 10^{-16} T_e^{0.45} \exp(-29.94/T_e)$& 28.08 \\ \hline
R19&\ce{H2^+ + e -> H+ + H + e} & $ 1.5 \times 10^{-13} \exp(-1.97/T_e)$ & 12.68 \\
\hline
R20&\ce{H3^+ + e -> 3H} & $2.8 \times 10^{-15} T_e^{0.48}$ & \\
\hline
R21&\ce{H3^+ + e -> H2 + H} & $1.6 \times 10^{-15} T_e^{0.48}$ & \\
\hline
R22&\ce{H2^+ + e -> 2H} & $1.4 \times 10^{-15} T_e^{0.43}$ & \\
\hline
R23&\ce{H2^+ + H2 -> H3^+ + H} & $20 \times 10^{-16}$ & \\
\hline
R24&\ce{e + H2 -> H2(v1) + e} & $3.086 \times 10^{-15} \exp(-0.714
+1.389\,(log(T_e))+
-0.897\,(log(T_e))^2
+0.206\,(log(T_e))^3
-0.00126\,(log(T_e))^4
-0.0175\,(log(T_e))^5
+0.000166\,(log(T_e))^6
+0.00200\,(log(T_e))^7
-0.000311\,(log(T_e))^8)$ & \\
\hline
R25&\ce{e + H2 -> H2(v2) + e} & $2.66 \times 10^{-16} \exp(-1.379
+2.630\,(log(T_e))+
-1.838\,(log(T_e))^2
+0.549\,(log(T_e))^3
-0.107\,(log(T_e))^4
+0.0966\,(log(T_e))^5
-0.0731\,(log(T_e))^6
+0.0210\,(log(T_e))^7
-0.00203\,(log(T_e))^8)$ & \\
\hline
R26&\ce{e + H2 -> H2(v3) + e} & $2.5 \times 10^{-17} \exp(-1.396
+2.937\,(log(T_e))+
-2.573\,(log(T_e))^2
+1.206\,(log(T_e))^3
-0.263\,(log(T_e))^4
-0.0185\,(log(T_e))^5
+0.0101\,(log(T_e))^6
+0.00147\,(log(T_e))^7
-0.00043\,(log(T_e))^8)$ & \\
\hline
\caption{Reactions, rate coefficients (final reaction rates in $\mathrm{\#/m^3/s}$) and electron energy loss, for \ce{H2} plasma chemistry. Electron temperature, $T_e$, is in units of eV for reactions R15-R23, while it is in K for reactions R24-26. R15-R23 rate coefficients are obtained from Sode et al. \cite{sode2013ion} while R24-26 rate coefficients are obtained from an offline Boltzmann solve using BOLSIG+ \cite{hagelaar2005solving}.}
\label{H2_rxns}
\end{longtable}

\begin{longtable}{|p{0.05\linewidth}|p{0.35\linewidth}|p{0.35\linewidth}|p{0.2\linewidth}|}
\hline
\hline
No.&Reaction& Rate Coefficient (K, m, \#)& Energy (eV) \\
\hline
\hline
R27&\ce{e + ArH^+ -> Ar + H}&$10^{-15}$& \\
\hline
R28&\ce{ArH^+ + H2 -> H3^{+} + Ar}&$6.3 \times 10^{-16}$&\\
\hline
R29&\ce{H2^+ + Ar -> ArH^+ + H}&$21.0 \times 10^{-16}$ & \\ \hline
R30&\ce{H2^+ + Ar -> Ar^+ + H2}&$2.0 \times 10^{-16}$& \\ \hline
R31&\ce{H3^+ + Ar -> ArH^+ + H2} & $3.7 \times 10^{-16}$ & \\
\hline
R32&\ce{Ar^+ + H2 -> ArH^+ + H} & $8.7 \times 10^{-16}$ & \\
\hline
R33&\ce{Ar^+ + H2 -> H2^+ + Ar} & $0.2 \times 10^{-16}$ & \\
\hline
\caption{Reactions, rate coefficients (final reaction rates in $\mathrm{\#/m^3/s}$) and electron energy loss, for \ce{Ar}-\ce{H2} plasma chemistry. These rates are obtained from Sode et al. \cite{sode2013ion}.}
\label{ArH2_rxns}
\end{longtable}
\begin{longtable}{|p{0.1\linewidth}|p{0.4\linewidth}|p{0.2\linewidth}|}
\hline
\hline
Species & Mobility ($\mathrm{m^2/V/s}$) & Ref. \\
\hline
\hline
\ce{e} & $\exp(0.05481-2.911 \tanh(0.289 \log(E/N) +0.543))(1.1296e25/N)$ & BOLSIG+ \cite{hagelaar2005solving} \\
\hline
\ce{Ar+} & $ 1.85 \times 10^{-4} \frac{N_0}{N}$ & \cite{iontransportIII}\\ \hline
\ce{Ar_2^+} & $1.835 \times 10^{-4} \frac{N_0}{N}$ & \cite{iontransportIII}\\ \hline
\ce{H+} & $30 \times 10^{-4} \frac{N_0}{N}$ &
\cite{iontransportI}\\ \hline
\ce{H_2+} & $25.0 \times 10^{-4} \frac{N_0}{N}$ & \cite{iontransportIII} \\ \hline
\ce{H_3+} & $32.0 \times 10^{-4} \frac{N_0}{N}$ & \cite{iontransportI} \\ 
\hline
\ce{ArH^+} & $19.0 \times 10^{-4} \frac{N_0}{N}$ & \cite{iontransportI} \\
\hline
\caption{Electron and ion mobility used in the simulations presented in this work. Here, $E/N$ denotes the reduced electric field in units of Td. $N$ is the background gas number density and $N_0=2.68e25$ is the gas number density at standard conditions (Loshmidt constant). The diffusion coefficients for electron and ionic species are calculated using the Einstein relation. The diffusion coefficients for neutral species are set the same as their corresponding ionic counterparts with the same molecular structure.}
\label{ArH2_mob}
\end{longtable}

\end{document}